%Paper: hep-th/9410167
%From: Chris Hull <C.M.Hull@qmw.ac.uk>
%Date: Fri, 21 Oct 94 17:53:51 +0100
%Date (revised): Fri, 13 Jan 95 18:34:49 GMT

%Dec 12 revised texfile

  %%%%%%%This requires the PHYZZX.TEX macropackage

%%%%%%%If you do not have the msbm fonts, delete the following 4 lines
\font\mybb=msbm10 at 12pt
\def\bb#1{\hbox{\mybb#1}}
\def\Z {\bb{Z}}
\def\R {\bb{R}}
%%%%%%%%%%%%
%%%and replace with the following 2 lines (without %)
%\def\Z {Z}
%\def\R {R}
%%%%%%%%%%

\tolerance=10000
\input phyzzx

 \def\unit{\hbox to 3.3pt{\hskip1.3pt \vrule height 7pt width .4pt \hskip.7pt
\vrule height 7.85pt width .4pt \kern-2.4pt
\hrulefill \kern-3pt
\raise 4pt\hbox{\char'40}}}
\def\II{{\unit}}
\def\cM {{\cal{M}}}
\def\half{{\textstyle {1 \over 2}}}

%%%%%%%%%%%%%%%%%%%%%%%%%%%%%%%%%%%%%%%%%%%%%%%%%%%%%%%%%%%%%%%%%%%%
\REF\TD {A. Giveon, M. Porrati and E. Rabinovici, Phys. Rep. {\bf 244}
(1994) 77.}
\REF\sdual{A. Font, L. Ibanez, D. Lust and F. Quevedo, Phys. Lett. {\bf B249}
(1990) 35; S.J. Rey, Phys. Rev. {\bf D43}  (1991) 526.}
\REF\SS {J.H. Schwarz and A. Sen, Nucl. Phys. {\bf B411} (1994) 35; Phys.
Lett. {\bf 312B} (1993) 105.}
\REF\Sen {A. Sen, Nucl. Phys. {\bf B404} (1993) 109; Phys. Lett. {\bf 303B}
(1993); Int. J. Mod. Phys. {\bf A8} (1993) 5079; Mod. Phys. Lett. {\bf
A8} (1993) 2023.}
\REF\Narain {K.S.  Narain, Phys. Lett. {\bf B169} (1986) 41.}
\REF\NSW {K.S.  Narain,  M.H. Sarmadi and E.Witten, Nucl. Phys. {\bf B279}
(1987) 369.}
\REF\CJ {E. Cremmer and B. Julia, Phys. Lett. {\bf 80B} (1978) 48; Nucl.
Phys. {\bf B159} (1979) 141.}
\REF\GH {G.W. Gibbons and C.M. Hull, Phys. Lett. {\bf 109B} (1982) 190.}
\REF\G {G.W. Gibbons, in {\it Supersymmetry, Supergravity and Related
Topics}, eds. F. del Aguila, J. A. de Azc{\' a}rraga and L.E. Iba{\~
n}ez, (World Scientific 1985).}
\REF\Kalort{
R. Kallosh and  T. Ortin, Phys.Rev. {\bf D48}  (1993) 742. }
\REF\GP {G.W. Gibbons and M.J. Perry, Nucl. Phys. {\bf B248} (1984) 629.}
\REF\bhstrings{S. W. Hawking, Monthly Notices Roy. Astron. Soc.
 {\bf 152} (1971) 75; Abdus Salam in
   {\it Quantum Gravity: an Oxford Symposium} (Eds. Isham, Penrose
and Sciama, O.U.P. 1975); G. t'Hooft, Nucl. Phys. {\bf B335} (1990) 138.}
\REF\bhstringsduff{
 M. J. Duff, R. R. Khuri, R. Minasian and J. Rahmfeld,  Nucl. Phys.
 {\bf B418} (1994) 195.}
\REF\DR{L. Susskind, preprint hep-th/9309145; J.G. Russo and L.
Susskind, preprint hep-th/9405117.}
\REF\Druff{ M.J. Duff and J. Rahmfeld, preprint
hep-th/9406105.}
\REF\HL {J. Harvey, J. Liu, Phys. Lett. {\bf B268} (1991) 40.}
\REF\GKLTT {G.W. Gibbons, D. Kastor, L. London, P.K. Townsend and J.
Traschen, Nucl. Phys. {\bf B416} (1994) 850.}
\REF\DGHR {A. Dabholkar, G.W. Gibbons, J.A. Harvey and F. Ruiz-Ruiz,
Nucl. Phys. {\bf B340} (1990) 33; M.J. Duff, G.W. Gibbons and P.K. Townsend,
Phys. Lett. {\bf 332B} (1994) 321.}
\REF\CHS {C. Callan, J. Harvey and A. Strominger, Nucl. Phys. {\bf B359}
(1991)
611.}
\REF\DL {M.J. Duff and J.X. Lu, Nucl. Phys. {\bf B354} (1991) 141;
Phys. Lett. {\bf 273B} (1991) 409.}
\REF\HS {G.T. Horowitz and A. Strominger, Nucl. Phys. {\bf B360} (1991)
197.}
\REF\DLb {M.J. Duff and J.X. Lu, Nucl. Phys. {\bf B416} (1993)  301.}
\REF\AGIT {J. A. de Azc{\' a}rraga, J.P. Gauntlett, J.M. Izquierdo and
P.K. Townsend, Phys. Rev. Lett. {\bf 63} (1989) 2443.}
\REF\Wone {E. Witten, Phys. Lett. {\bf 153B} (1985) 243.}
\REF\PKTGSW {P.K. Townsend, Phys. Lett. {\bf 139B} (1984) 283; M.B.
Green, J.H. Schwarz and P.C. West, Nucl. Phys. {\bf B254} (1985) 327.}
\REF\BST {E. Bergshoeff, E. Sezgin and P.K. Townsend, Phys. Lett. {\bf 189B}
(1987) 75; Ann. Phys. (N.Y.) {\bf 185} (1988) 330.}
\REF\CJS {E. Cremmer, B. Julia and J. Scherk, Phys. Lett. {\bf 76B} (1978)
409.}
\REF\DNP {M.J. Duff, B.E.W. Nilsson and C.N. Pope, Phys. Lett {\bf 129}B
(1983) 39.}
\REF\DS {M.J. Duff and K.S. Stelle, Phys. Lett. {\bf 253B} (1991) 113.}
\REF\Gu {R. G{\" u}ven, Phys. Lett. {\bf 276B} (1992) 49.}
\REF\GZ {M.K. Gaillard and B. Zumino, Nucl. Phys. {\bf B193} (1981) 221.}
\REF\GST{M. G{\" u}naydin, G. Sierra and P.K. Townsend, Phys. Lett. {\bf
133B}
(1983) 72;  Nucl. Phys. {\bf B242} (1984) 244.}
\REF\Wtwo {E. Witten, Phys. Lett. {\bf 86B} (1979) 283.}
\REF\WO {E. Witten and D. Olive, Phys. Lett. {\bf B78} (1978) 97.}
\REF\cmh{C.M. Hull, Ph.D. Thesis, Cambridge 1983.}
\REF\GETC {G.W. Gibbons, Nucl. Phys. {\bf B207} (1982) 337; G.W. Gibbons and
K. Maeda, Nucl. Phys. {\bf B298} (1988) 741; D. Garfinkle, G.T. Horowitz and
A. Strominger, Phys. Rev. {\bf D43}, (1991) 3140; C.F. Holzhey and F. Wilczek,
Nucl. Phys. {\bf B380} (1992) 447; R. Kallosh, A. Linde, T. Ortin, A. Peet, A.
Van
Proeyen, Phys.Rev. {\bf D46} (992) 5278.}
\REF\DKMR {M.J. Duff, R.R. Khuri, R. Minasian and J. Rahmfeld, Nucl. Phys. {\bf
B418}
 (1994) 195.}
\REF\GHT {G.W. Gibbons, G.T. Horowitz and P.K. Townsend, {\it
Higher-dimensional
resolution of dilatonic black hole singularities}, Class. Quantum Grav. {\it in
press}.}
\REF\Gauntlett {J. Gauntlett, Nucl. Phys. {\bf B400} (1993) 103.}
\REF\Ruback {P. Ruback, Commun. Math. Phys. 107 (1986) 93.}
\REF\Shiraishi {K. Shiraishi, J. Math. Phys. {\bf 34} (4) (1993) 1480.}
\REF\DLtwo {M.J. Duff and J.X. Lu, Nucl. Phys. {\bf B390} (1993) 276.}
\REF\FS {A.G. Felce and T.M. Samols, Phys. Lett. {\bf B308} (1993) 30.}
\REF\GD {Bluhm, P. Goddard and L. Dolan, Nucl. Phys.
 {\bf B289}  (1987), 364; {\bf B309} (1988) 330. }
\REF\maha{J. Maharana and J.H. Schwarz, Nucl. Phys. {\bf B390} (1993) 3.}
\REF\dR {M. de Roo, Nucl. Phys. {\bf B255} (1985) 515.}
\REF\GHL {J. Gauntlett, J. Harvey and J. Liu, Nucl. Phys. {\bf B409}
(1993) 363; J. Gauntlett and J. Harvey, {\it S-Duality and the spectrum of
magnetic monopoles in heterotic string theory}, preprint EFI-94-36.}
\REF\SGP {R. Sorkin, Phys. Rev. Lett. {\bf 51} (1983) 87 ; D. Gross and
M. Perry, Nucl. Phys. {\bf B226} (1983) 29.}
\REF\khuria{R.R. Khuri,
Phys. Lett. {\bf  B259} (1991) 261.}
\REF\khurib{R.R. Khuri, Nucl. Phys. {\bf B387} (1992) 315.}
\REF\fish{T. Banks, M. Dine, H. Dijkstra and W. Fischler, Phys. Lett. {\bf
B 212} (1988) 45.}
\REF\prep{C.M. Hull, in preparation.}
\REF\RNEP {R. Nepomechie, Phys. Rev. {\bf D31} (1985) 1921.}
\REF\RNEPA { A. Strominger, Nucl. Phys. {\bf B343} (1990) 167.}
\REF\DLPRL { M.J. Duff and J.X. Lu, Phys. Rev. Lett. {\bf 66} (1991) 1402.}
\REF\hypkah{C.M. Hull, Nucl. Phys. {\bf B260} (1985) 182 ;
L. Alvarez-Gaum\' e and P. Ginsparg, Commun. Math. Phys. {\bf 102} (1985) 311.}
\REF\hortset{G.T. Horowitz and A.A. Tseytlin, I.C. preprints Imperial
/TP/93-94/38,
Imperial /TP/93-94/54.}
\REF\ABEQUIV {J. Dai, R.G. Leigh and J. Polchinski, Mod. Phys. Lett. {\bf
A4} (1989) 2073; M. Dine, P. Huet and N. Seiberg, Nucl. Phys. {\bf B322}
(1989) 301.}
\REF\hypbps{ M. Atiyah and N. Hitchin, Phys. Lett. {\bf 107A} (1985) 21; Phil.
Trans.
R. Soc. Lond. {\bf A315} (1985) 459; {\it The Geometry and Dynamics of Magnetic
Monopoles}, Princeton University Press (1988).}
\REF\julia {B. Julia in {\it Supergravity and Superspace}, S.W. Hawking and M.
Rocek, C.U.P.
Cambridge,  1981. }
\REF\juliab{B. Julia in {\it Lectures in Applied Mathematics}, AMS,  vol. 21,
355, 1985. }
\REF\DLmem{M.J. Duff and J.X. Lu, Nucl. Phys. {\bf B347} (1990) 394.}
\REF\Nic{R.W. Gebert and  H. Nicolai, DESY-94-106,   hep-th
9406175.}
\REF\Sezsal{Supergravity Theories in Diverse Dimensions, A. Salam and E. Sezgin
(eds.),
World Scientific, Singapore.}
\REF\senthree {A. Sen, Tata Inst. preprint TIFR-TH-94-19, hep-th 9408083.}
\REF\Cremmer{E.Cremmer, in {\it Supergravity and Superspace}, S.W. Hawking and
M. Rocek, C.U.P.
Cambridge,  1981.}
\REF\Nepteit{ C. Teitelboim,
Phys. Lett. {\bf  B67} (1986) 63 and 69.}
\REF\AETW {A. Ach\'ucarro, J. Evans, P.K. Townsend and D. Wiltshire,
Phys. Lett. {\bf 198B} (1987) 441.}
\REF\cmhpp{C.M. Hull,  Phys. Lett. 139B (1984) 39.}
\REF\DGT {M.J. Duff, G.W. Gibbons and P.K. Townsend, Phys. Lett. {\bf 332
B} (1994) 321.}
\REF\SenSegal {A. Sen,  Phys.Lett. {\bf B329}, (1994) 217.}
\REF\VW {C. Vafa and E. Witten, Harvard preprint HUTP-94-A017,
hep-th/9408074.}
\REF\PKT {P.K. Townsend, Phys. Lett. {\bf 202B} (1988) 53.}

%%%%%%%%%%%%%%%%%%%%%%%%%%%%%%%%%%%%%%%%%%%%%%%%%%%%%%%%%%%%%%%%%%%%

\Pubnum{ \vbox{ \hbox {QMW-94-30} \hbox{R/94/33}  \hbox{hep-th/9410167}} }
\pubtype{}
\date{October, 1994}

\titlepage

\title {\bf  Unity of Superstring Dualities}

\author{C.M. Hull}
\address{Physics Department,
Queen Mary and Westfield College,
\break
Mile End Road, London E1 4NS, U.K.}
\andauthor{P. K. Townsend}
\address{DAMTP, University of Cambridge,
\break
Silver Street, Cambridge CB3 9EW,  U.K.}
\vskip 0.5 cm
%\centerline{{\bf 19th October DRAFT}}
\vskip 0.5cm

\abstract {The effective action for type II string theory compactified
on a six torus is $N=8$ supergravity, which is known to have an $E_{7}$ duality
symmetry. We show that this is broken by quantum effects to a discrete
subgroup, $E_7(\Z)$, which contains both the T-duality group $O(6,6;\Z)$ and
the S-duality group $SL(2;\Z)$. We present evidence for the conjecture that
$E_7(\Z)$ is an exact \lq U-duality' symmetry of type II string theory. This
conjecture requires certain extreme black hole states to be identified with
massive modes of the fundamental string. The gauge bosons from the
Ramond-Ramond sector couple not to string excitations but to solitons.
We discuss similar issues in the context of toroidal string compactifications
to
other dimensions, compactifications of the type II string on $K_3\times T^2$
and compactifications of eleven-dimensional supermembrane theory.}

\endpage

%%%%%%%%%%%%%%%%%%%%%%%%%%%%%%%%
%
% S-Tables Macro
%
\message{S-Tables Macro v1.0, ACS, TAMU (RANHELP@VENUS.TAMU.EDU)}
%
% Help Text
%
\newhelp\stablestylehelp{You must choose a style between 0 and 3.}%
\newhelp\stablelinehelp{You should not use special hrules when stretching
a table.}%
\newhelp\stablesmultiplehelp{You have tried to place an S-Table inside another
S-Table.  I would recommend not going on.}%
%
% Line Thicknesses (Values)
%
\newdimen\stablesthinline
\stablesthinline=0.4pt
\newdimen\stablesthickline
\stablesthickline=1pt
%
% Border and Internal Line Thicknesses
%
\newif\ifstablesborderthin
\stablesborderthinfalse
\newif\ifstablesinternalthin
\stablesinternalthintrue
\newif\ifstablesomit
\newif\ifstablemode
\newif\ifstablesright
\stablesrightfalse
%
% Save Registers
%
\newdimen\stablesbaselineskip
\newdimen\stableslineskip
\newdimen\stableslineskiplimit
%
% Counters
%
\newcount\stablesmode
\newcount\stableslines
\newcount\stablestemp
\stablestemp=3
\newcount\stablescount
\stablescount=0
\newcount\stableslinet
\stableslinet=0
%
% Table Style Selection
%
% 0 - Centered
% 1 - Left Justified
% 2 - Right Justified
% 3 - Not Justified
%
\newcount\stablestyle
\stablestyle=0
%
% Element Buffering Definitions
%
\def\stablesleft{\quad\hfil}%
\def\stablesright{\hfil\quad}%
%
% Vertical Bar Activation
%
\catcode`\|=\active%
%
% Strut Control
%
\newcount\stablestrutsize
\newbox\stablestrutbox
\setbox\stablestrutbox=\hbox{\vrule height10pt depth5pt width0pt}
\def\stablestrut{\relax\ifmmode%
                         \copy\stablestrutbox%
                       \else%
                         \unhcopy\stablestrutbox%
                       \fi}%
%
% Misc. Internal Stuff
%
\newdimen\stablesborderwidth
\newdimen\stablesinternalwidth
\newdimen\stablesdummy
\newcount\stablesdummyc
\newif\ifstablesin
\stablesinfalse
%
% Table Macros
%
\def\begintable{\stablestart%
  \stablemodetrue%
  \stablesadj%
  \halign%
  \stablesdef}%
\def\stablesadj{%
  \ifcase\stablestyle%
    \hbox to \hsize\bgroup\hss\vbox\bgroup%
  \or%
    \hbox to \hsize\bgroup\vbox\bgroup%
  \or%
    \hbox to \hsize\bgroup\hss\vbox\bgroup%
  \or%
    \hbox\bgroup\vbox\bgroup%
  \else%
    \errhelp=\stablestylehelp%
    \errmessage{Invalid style selected, using default}%
    \hbox to \hsize\bgroup\hss\vbox\bgroup%
  \fi}%
\def\stablesend{\egroup%
  \ifcase\stablestyle%
    \hss\egroup%
  \or%
    \hss\egroup%
  \or%
    \egroup%
  \or%
    \egroup%
  \else%
    \hss\egroup%
  \fi}%
\def\stablestart{%
  \ifstablesin%
    \errhelp=\stablesmultiplehelp%
    \errmessage{An S-Table cannot be placed within an S-Table!}%
  \fi
  \global\stablesintrue%
  \global\advance\stablescount by 1%
  \message{<S-Tables Generating Table \number\stablescount}%
  \begingroup%
  \stablestrutsize=\ht\stablestrutbox%
  \advance\stablestrutsize by \dp\stablestrutbox%
  \ifstablesborderthin%
    \stablesborderwidth=\stablesthinline%
  \else%
    \stablesborderwidth=\stablesthickline%
  \fi%
  \ifstablesinternalthin%
    \stablesinternalwidth=\stablesthinline%
  \else%
    \stablesinternalwidth=\stablesthickline%
  \fi%
  \tabskip=0pt%
  \stablesbaselineskip=\baselineskip%
  \stableslineskip=\lineskip%
  \stableslineskiplimit=\lineskiplimit%
  \offinterlineskip%
  \def\borderrule{\vrule width \stablesborderwidth}%
  \def\internalrule{\vrule width \stablesinternalwidth}%
  \def\thinline{\noalign{\hrule height \stablesthinline}}%
  \def\thickline{\noalign{\hrule height \stablesthickline}}%
  \def\trule{\omit\leaders\hrule height \stablesthinline\hfill}%
  \def\ttrule{\omit\leaders\hrule height \stablesthickline\hfill}%
  \def\tttrule##1{\omit\leaders\hrule height ##1\hfill}%
  \def\stablesel{&\omit\global\stablesmode=0%
    \global\advance\stableslines by 1\borderrule\hfil\cr}%
  \def\el{\stablesel&}%
  \def\elt{\stablesel\thinline&}%
  \def\eltt{\stablesel\thickline&}%
  \def\elttt##1{\stablesel\noalign{\hrule height ##1}&}%
  \def\elspec{&\omit\hfil\borderrule\cr\omit\borderrule&%
              \ifstablemode%
              \else%
                \errhelp=\stablelinehelp%
                \errmessage{Special ruling will not display properly}%
              \fi}%
  \def\stmultispan##1{\mscount=##1 \loop\ifnum\mscount>3 \stspan\repeat}%
  \def\stspan{\span\omit \advance\mscount by -1}%
  \def\multicolumn##1{\omit\multiply\stablestemp by ##1%
     \stmultispan{\stablestemp}%
     \advance\stablesmode by ##1%
     \advance\stablesmode by -1%
     \stablestemp=3}%
  \def\multirow##1{\stablesdummyc=##1\parindent=0pt\setbox0\hbox\bgroup%
    \aftergroup\emultirow\let\temp=}
  \def\emultirow{\setbox1\vbox to\stablesdummyc\stablestrutsize%
    {\hsize\wd0\vfil\box0\vfil}%
    \ht1=\ht\stablestrutbox%
    \dp1=\dp\stablestrutbox%
    \box1}%
  \def\stpar##1{\vtop\bgroup\hsize ##1%
     \baselineskip=\stablesbaselineskip%
     \lineskip=\stableslineskip%
     \lineskiplimit=\stableslineskiplimit\bgroup\aftergroup\estpar\let\temp=}%
  \def\estpar{\vskip 6pt\egroup}%
  \def\stparrow##1##2{\stablesdummy=##2%
     \setbox0=\vtop to ##1\stablestrutsize\bgroup%
     \hsize\stablesdummy%
     \baselineskip=\stablesbaselineskip%
     \lineskip=\stableslineskip%
     \lineskiplimit=\stableslineskiplimit%
     \bgroup\vfil\aftergroup\estparrow%
     \let\temp=}%
  \def\estparrow{\vfil\egroup%
     \ht0=\ht\stablestrutbox%
     \dp0=\dp\stablestrutbox%
     \wd0=\stablesdummy%
     \box0}%
  \def|{\global\advance\stablesmode by 1&&&}%
  \def\|{\global\advance\stablesmode by 1&\omit\vrule width 0pt%
         \hfil&&}%
  \def\vt{\global\advance\stablesmode by 1&\omit\vrule width \stablesthinline%
          \hfil&&}%
  \def\vtt{\global\advance\stablesmode by 1&\omit\vrule width
\stablesthickline%
          \hfil&&}%
  \def\vttt##1{\global\advance\stablesmode by 1&\omit\vrule width ##1%
          \hfil&&}%
  \def\vtr{\global\advance\stablesmode by 1&\omit\hfil\vrule width%
           \stablesthinline&&}%
  \def\vttr{\global\advance\stablesmode by 1&\omit\hfil\vrule width%
            \stablesthickline&&}%
  \def\vtttr##1{\global\advance\stablesmode by 1&\omit\hfil\vrule width ##1&&}%
  \stableslines=0%
  \stablesomitfalse}
\def\stablesdef{\bgroup\stablestrut\borderrule##\tabskip=0pt plus 1fil%
  &\stablesleft##\stablesright%
  &##\ifstablesright\hfill\fi\internalrule\ifstablesright\else\hfill\fi%
  \tabskip 0pt&&##\hfil\tabskip=0pt plus 1fil%
  &\stablesleft##\stablesright%
  &##\ifstablesright\hfill\fi\internalrule\ifstablesright\else\hfill\fi%
  \tabskip=0pt\cr%
  \ifstablesborderthin%
    \thinline%
  \else%
    \thickline%
  \fi&%
}%
\def\endtable{\advance\stableslines by 1\advance\stablesmode by 1%
   \message{- Rows: \number\stableslines, Columns:  \number\stablesmode>}%
   \stablesel%
   \ifstablesborderthin%
     \thinline%
   \else%
     \thickline%
   \fi%
   \egroup\stablesend%
\endgroup%
\global\stablesinfalse}
%
% end of STABLES.TEX
%

%%%%%%%%%%%%%%%%%%%%%%%%%%%%%%%INTRODUCTION%%%%%%%%%%%%%%%%%%%%%%%%%%%%%%
%%

\chapter{ Introduction}

String theory in a given background can be formulated in terms of a sum over
world-sheet fields, (super-) moduli  and topologies of a world-sheet
sigma-model with the background spacetime as its target space. Different
backgrounds may define the same quantum string theory, however, in which case
they must be identified. The transformations between equivalent
backgrounds generally define a discrete group and such discrete gauge
symmetries are referred to as duality symmetries of the string theory.  An
example is T-duality, which relates spacetime geometries possessing a compact
abelian isometry group (see [\TD] and references therein). The simplest case
arises from compactification of the string theory on a circle since a circle
of radius $R$ defines the same two-dimensional quantum field theory, and hence
the same string theory, as that on a circle of radius $\alpha'/R$.
T-dualities are non-perturbative in the sigma-model coupling constant
$\alpha'$ but valid order by order in the string coupling constant $g$. Some
string theories may have additional discrete symmetries which are
perturbative in $\alpha'$ but non-perturbative in $g$. An example is the
conjectured S-duality of the heterotic string compactified on a six-torus
[\sdual,\SS,\Sen]. In this paper we investigate duality symmetries of the type
II string compactified to four dimensions and present evidence for a new
`U-duality' symmetry which unifies the S and T dualities and mixes sigma-model
and string coupling constants.

Consider a compactified string for which the internal space is an
$n$-torus with constant metric $g_{ij}$ and antisymmetric tensor $b_{ij}$.
The low-energy effective field theory includes a spacetime sigma model whose
target space is the moduli space $O(n,n)/\big[O(n)\times
O(n)\big]$ of the torus, and  the constants $g_{ij}$ and $b_{ij}$ are
the expectation values of the $n^2$ scalar fields. There is a natural
action of $O(n,n)$ on the moduli space. In general this takes one string
theory into a different one but a discrete $O(n,n;\Z)$ subgroup takes a
given string theory into an equivalent one. This is the T-duality
group of the toroidally-compactified string and the true moduli space of the
string theory is the moduli space of the torus factored by the discrete
T-duality group. There is a generalization to Narain compactifications on the
`(p,q)-torus' $T(p,q)$ for which the left-moving modes of the string are
compactified on a $p$-torus and the right-moving ones on a $q$-torus
[\Narain]. In this case the moduli space is $O(p,q)/\big[O(p)\times
O(q)\big]$ factored by the T-duality group $O(p,q;\Z)$. The $T(6,22)$ case
is relevant to the heterotic string compactified to four dimensions which has
$O(6,22;\Z)$ as its T-duality group. At a generic point in the moduli space
the effective field theory is $N=4$ supergravity coupled to 22 abelian vector
multiplets, giving a total of 28 abelian vector gauge fields [\NSW] with
gauge group $U(1)^{28}$. It follows from the compactness of the full gauge
group for all 28 vector gauge fields that any electric or magnetic charges
are quantized. The effective field theory has an $SL(2;\R)\times O(6,22)$
invariance of the equations of motion which, due to  the charge quantization
and the fact that states carrying all types of charge can be found in the
spectrum, is broken to the discrete subgroup $SL(2;\Z)\times O(6,22;\Z)$.
The $O(6,22;\Z)$ factor extends to the T-duality group of the full string
theory. It has been conjectured that the $SL(2;\Z)$ factor also extends to a
symmetry of the full string theory [\SS]. This is the S-duality group of the
heterotic string. It acts on the dilaton field $\Phi$ and the axion field
$\psi$ (obtained by dualizing the four-dimensional two-form gauge field
$b_{\mu\nu}$ that couples to the string) via fractional linear transformations
of the complex scalar $\psi +ie^{-\Phi}$ and on the abelian field strengths by
a generalized electric-magnetic duality. One of the $SL(2;\Z)$ transformations
interchanges the electric and magnetic fields  and, when $\psi=0$, takes
$\Phi$ to $-\Phi$ which, since the expectation value of $e^\Phi$ can be
identified with the string coupling constant $g$, takes $g$ to $1/g$, and so
interchanges strong and weak coupling.

Consider now the compactification of the type IIA or type IIB superstring
to four dimensions on a six-torus. The low-energy effective field theory is
$N=8$ supergravity [\CJ], which has 28 abelian vector gauge fields and 70
scalar
fields taking values in $E_{7(7)}/\big[ SU(8)/Z_2\big]$. The equations of
motion are invariant under the action of $E_{7(7)}$ [\CJ], which contains
$SL(2;\R)\times O(6,6)$ as a maximal subgroup. We shall show that
certain quantum mechanical effects break $E_{7(7)}$ to a discrete subgroup
which we shall call $E_7(\Z)$, and this implies a breaking of the maximal
$SL(2;\R)\times O(6,6)$ subgroup to $SL(2;\Z)\times O(6,6;\Z)$. The
$O(6,6;\Z)$ factor extends to the full string theory as the T-duality group
and it is natural to conjecture that the $SL(2;\Z)$ factor also extends to the
full string theory as an S-duality group. In fact, we shall present evidence
for the much stronger conjecture that the full $E_7(\Z)$ group (to be defined
below) extends to the full string theory as a new unified duality group,
which we call U-duality. U-duality acts on the abelian gauge fields through a
generalized electromagnetic duality and on the 70 scalar fields,  the constant
parts of which can each be thought of as a coupling constant of the theory. The
zero-mode of the dilaton is related to the string coupling $g$, while $21$ of
the
scalar zero-modes are the moduli of the metric on the 6-torus, and the others
parameterise
the space of constant anti-symmetric tensor gauge fields on the six-torus.
U-duality implies that all 70 coupling constants are on a similar footing
despite the fact that the standard perturbative formulation of string theory
assigns a special r\^ ole to one of them.

Whereas T-duality is known to be an exact symmetry of string theory at each
order in the string coupling constant $g$, the conjectured S-duality and
U-duality are non-perturbative and so cannot be established within
a perturbative formulation of string theory. However, it was pointed out in
[\Sen] in the context of the heterotic string that there are a number of
quantities for which the tree level results are known to be, or believed to be,
exact, allowing a check on S-duality by a perturbative, or semi-classical,
calculation. We shall show that U-duality for the type II string
passes the same tests.

First, for compactifications of the type II string that preserve at least $N=4$
supersymmetry, the low energy effective field theory for the massless modes is
a supergravity theory whose form is determined uniquely by its local symmetries
and is therefore not changed by quantum corrections. Duality of the string
theory therefore implies the duality invariance of the equations of motion of
the supergravity theory. This prediction is easily checked because the
symmetries of the $N\ge4$ supergravity/matter theories have been known for some
time. In particular, the equations of motion of $N=8$ supergravity are
U-duality invariant. Another quantity that should be, and is, duality invariant
is the set of values of electric and magnetic charges allowed by the
Dirac-Schwinger-Zwanziger quantization condition. Third, the masses of states
carrying electric or magnetic charges satisfy a Bogomolnyi bound which, for the
compactifications considered here,
is believed to be unrenormalized to arbitrary order in the string coupling
constant. Duality invariance of the string theory requires this bound to be
duality invariant. For soliton states the Bogomolnyi bound can be found from a
classical bound on field configurations of the effective supergravity theory
carrying electric or magnetic charges that generalizes the bound obtained in
[\GH] for Maxwell-Einstein theory. We present this bound for $N=8$ supergravity
and show that it is U-duality invariant. Fourth, the spectrum of `Bogomolnyi
states'
saturating the Bogomolnyi bound  should also be
duality invariant. These states include winding and momentum modes of the
fundamental string and those found from quantization of solitons. We
shall assume that soliton solutions of the type II string can be identified
with
 those of its
effective $N=8$  supergravity theory, and these, as we shall see,
 include various types of extreme black
hole\foot{See [\G] for a discussion of the interpretation of extreme black
holes as solitons}.

One of the main concerns of this paper will be the Bogomolnyi states  of the
type II string theory
that break half the supersymmetry. As we shall see, the soliton states in this
category arise from
quantization of a particular class of extreme black hole solution of $N=8$
supergravity. It is essentially
automatic that all soliton states of the type II string fall into
representations of  the U-duality group because this is a symmetry
of the equations of motion of which the solitons are solutions\foot{They also
fall into supermultiplets because of the fermion zero modes in the presence of
an extreme black hole [\GP]}. A similar argument can be made for solitons of
the heterotic string; for example, extreme black hole solutions of the
low-energy field theory corresponding to the heterotic string fit into
$SL(2,\Z)$ representations [\Kalort].
There are two points to bear in mind, however.
First, a duality transformation not only produces new soliton solutions
from old but also changes the
vacuum, as the vacuum is parameterised  by
the scalar  expectation values and these change under duality.
We shall assume, as in [\Sen], that the new soliton state in the new
vacuum can be continued back to give a new soliton state in the old vacuum with
duality transformed electric and magnetic charges;
this is certainly possible at the level of solutions of the low energy
effective action, since the extreme black hole solutions
depend analytically on the scalar expectation values.
Combining U-duality transformations with analytic continuations of the scalar
field zero-modes in this way gives an $E_7(\Z) $ invariance of the spectrum of
soliton states in a given vacuum. (Note that whereas U-duality preserves
masses, combining this with a scalar zero-mode continuation gives a
transformation which changes masses and so is  not an invariance of the
Hamiltonian.) Second, the four-dimensional metrics of many extreme black hole
solitons are only
defined up to a conformal rescaling by the exponential of a scalar field
function that vanishes at spatial infinity. While the `Einstein' metric is
duality invariant, other metrics in the same conformal equivalence class will
not be. In general one should therefore think of duality as acting on conformal
equivalence classes of metrics and the issue arises as to which metric within
this class is the physically relevant one.
As we shall see, for the solutions considered here each conformal class of
metrics contains one that is (i) either completely regular or regular outside
and on an event horizon and (ii) such that its spatial sections
interpolate between topologically distinct vacua.
The extreme black hole solutions corresponding to these metrics might
reasonably be interpreted as solitons of the theory.

We now encounter an apparent contradiction with U-duality, and with S-duality,
of the type II string theory because the fundamental string excitations include
additional Bogomolnyi states which apparently cannot be assigned to duality
multiplets containing solitons because the soliton multiplets are already
complete. The only escape from this contradiction is to make the hypothesis
that the fundamental string states have already been counted among the soliton
states. In order for this to be possible there must be soliton states carrying
exactly the same quantum numbers as the fundamental Bogomolnyi states. This is
indeed the case. The idea that particles with masses larger than the Planck
mass, and hence a Compton wavelength less than their Schwarzschild radius,
should be regarded as black holes is an old one [\bhstrings,\bhstringsduff],
and it has recently been argued that Bogomolnyi states in the excitation
spectrum of the heterotic string should be identified with extreme
electrically-charged dilaton black holes [\DR,\Druff].
For the heterotic string, approximate solutions of the low-energy effective
action include
extreme black holes and self-gravitating BPS
monopoles [\HL,\GKLTT], and it is believed that these correspond to Bogomolnyi
solitons of the
heterotic string [\Sen]. Any magnetically charged soliton will have an
electrically
charged soliton partner generated by  the action of the $\Z _2$ electromagnetic
duality subgroup of
S-duality. Now, if the full string theory is S-duality invariant,   and this
$\Z _2$ subgroup  acts
on an electrically charged fundamental string state  to give a magnetically
charged soliton, as
argued for the heterotic string in [\Sen], then this fundamental string state
must be identified
with the corresponding electrically charged soliton.
 We shall return to these points later but it is worth noting here that
solitons of the low-energy effective $N=4$ or $N=8$ supergravity theory fit
into
representations of the $S \times T$ or $U$ duality as these are symmetries of
the supergravity equations of motion, and this is true {\it irrespective of
whether the duality symmetry is actually  a symmetry of the full heterotic or
type II string
theory}.

For compactifications of ten-dimensional string theories one expects solitons
of
the effective four-dimensional theory to have a ten-dimensional origin. For the
type II string we are able to identify the four dimensional solitons that break
half the supersymmetry of $N=8$ supergravity as six-torus `compactifications'
of
the extreme black $p$-branes of either IIA or IIB ten-dimensional supergravity
[\DGHR,\CHS,\DL,\HS,\DLb]. We note that, in this context, the Bogomolnyi bound
satisfied by these states can be seen to arise from the algebra of Noether
charges of the effective world-volume action [\AGIT]. Remarkably, the
solitonic states that are required to be identified with fundamental string
states are precisely those which have their ten-dimensional origin in the
string soliton or extreme black 1-brane solution, which couples to the same
two-form gauge field as the fundamental string. This suggests that we should
identify the fundamental ten dimensional string with the solitonic string. This
is consistent with a suggestion made in [\Wone], for other reasons, that the
four-dimensional heterotic string be identified with an axion string.

A similar analysis can be carried out for non-toroidal compactification.
A particularly interesting example is compactification of the type II
superstring on $K_3\times T^2$ [\PKTGSW] for which the effective
four-dimensional field theory turns out to be identical to the effective field
theory of the $T^6$-compactified heterotic string, and in particular has the
same $SL(2,\Z) \times O(6,22;\Z)$ duality group. Furthermore, the spectrum of
extreme black hole states is also the same. This raises the possibility that
the
two string theories might be {\it non-perturbatively} equivalent, even though
they differ perturbatively. Such an equivalence would clearly have significant
implications.

Finally we consider similar issues in the context of the 11-dimensional
supermembrane [\BST]. This couples naturally to 11-dimensional supergravity
[\CJS] and hence to $N=8$ supergravity after compactification on $T^7$ and
to $N=4$ supergravity coupled to 22 vector multiplets after compactification
on $K_3\times T^3$ [\DNP]. At present it is not known how to make sense of a
quantum supermembrane, so there is little understanding of what the massive
excitations might be. However, some progress can be made using the methods
sketched above for the string. We shall show that, if the elementary
supermembrane is identified with the solitonic membrane solution [\DS] of
11-dimensional supergravity and account is taken of the solitonic fivebrane
solution [\Gu], the results of this analysis for the four-dimensional theory
are exactly the same as those of the type II string.

%%%%%%%%%%%%%%%%%%%%%%%%%%%%%%%Chapter2%%%%%%%%%%%%%%%%%%%%%%%%%%%%%%%%

\chapter{ Charge Quantization and the Bogomolnyi Bound}

Consider the four-dimensional Lagrangian
$$
L=\sqrt
{-g} \left({1\over4}  R -\half g_{ij}(\phi)\partial_\mu\phi^i\partial^\mu\phi^j
-{1\over4}m_{IJ}(\phi) F^{\mu\nu}F_{\mu\nu}^J - {1\over8}
\varepsilon^{\mu\nu\rho\sigma}a_{IJ}(\phi)F_{\mu\nu}^I F_{\rho\sigma}^J\right)
\eqn\aone
$$
for space-time 4-metric $g_{\mu\nu}$, scalars $\phi^i$ taking values in a
sigma-model target space $\cM$ with metric $g_{ij}(\phi)$, and $k$ abelian
vector fields  $A_\mu^I$ with field strengths $F_{\mu\nu}^I$. The scalar
functions $m_{IJ}+ia_{IJ}$ are entries of a positive definite $k\times k$
Hermitian matrix. The bosonic sector of all supergravity theories without
scalar potentials or non-abelian gauge fields can be put in this form. We
shall be interested in those cases for which the equations of motion are
invariant under some symmetry group $G$, which is necessarily a subgroup of
$Sp(2k;\R)$ [\GZ] and an isometry group of $\cM$. Of principal interest here
are the special cases for which $\cM$ is the homogeneous space $G/H$ where $H$
is the maximal compact subgroup of $G$. These cases include many supergravity
theories, and all those with $N\ge4$ supersymmetry. For $N=4$ supergravity
coupled to $m$ vector multiplets $k=6+m$, $G= SL(2;\R)\times O(6,m)$ and $H=
U(1)\times O(6)\times O(m)$. For $N=8$ supergravity $k=28$, $G= E_{7(7)}$ and
$H=SU(8)$. For the `exceptional' $N=2$ supergravity [\GST], $k=28$,
$G=E_{7(-25)}$ and
$H=E_6\times U(1)$.

Defining
$$
G_{\mu\nu}{}_I = m_{IJ}{ \star F}_{\mu\nu}^J + a_{IJ}F_{\mu\nu}^J
\eqn\atwo
$$
where $\star F_{\mu\nu}^I={1\over2}\varepsilon_{\mu\nu\rho\sigma}
F^{\rho\sigma}{}^I$, the $A_\mu^I$ field equations and Bianchi identities
can be written in terms of the the $2k$-vector of two-forms
$$
{\cal F} = \pmatrix{ F^I\cr G_I}
\eqn\athree
$$
as simply $d{\cal F}=0$. The group $G$ acts on the scalars through isometries
of $\cM$ and on  ${\cal F}$ as ${\cal F}
\rightarrow \Lambda {\cal F}$ where $\Lambda\in G  \subseteq Sp(2k;\R)$ is a
$2k\times 2k$ matrix preserving the $2k\times 2k$ matrix
$$
\Omega = \pmatrix{0& \II \cr -\II &0}
\eqn\afour
$$
An alternative way to represent the $G/H$ sigma model is in terms of a
$G$-valued field $V(x)$ which transforms under rigid $G$-transformations by
right multiplication and under local $H$ transformations by left
multiplication
$$
V(x)\rightarrow h(x)V(x)\Lambda^{-1} \qquad h\in H\ ,\ \Lambda\in G
\eqn\afive
$$
The local $H$-invariance can be used to set $V\in G/H$. Note that
$\bar {\cal F} \equiv V{\cal F}$ is $G$-invariant. In most cases of interest,
the scalar coset space can be parameterised by the complex scalars
$z_{IJ}=a_{IJ}+im_{IJ}$ which take values in a generalised upper-half-plane
($m_{IJ}$ is positive definite) and the group $G$ acts on $z_{IJ}$ by
fractional linear transformations.  (This can be seen for  $N=8$ supergravity
as follows.
In the symmetric gauge [\CJ], the coset is parameterised by a scalar $y_{IJ}$
which transforms under fractional linear transformations under $G$.  However,
$z_{IJ}$ is related to $y_{IJ}$
by a fractional linear transformation, $z_{IJ}= i ( \II + \bar y) /( \II - \bar
y)$, so that $z$ in turn transforms under
$G$ by fractional linear transformations. Similar results follow for $N<8$
supergravities by truncation.)

We now define the charges
$$
Q^I =\oint_\Sigma \star F^I \qquad p_I =  {1 \over 2\pi} \oint_\Sigma F^I
\qquad q_I =
\oint_\Sigma G_I
\eqn\asix
$$
as integrals of two-forms over a two-sphere $\Sigma$ at spatial infinity. The
charges $p^I$ and $q_I$ are the magnetic charges and the Noether electric
charges, respectively. The charges $Q^I$ are the electric charges describing
the
$1/r^2$ fall-off of the radial components of the electric fields, $F^I_{0r}$,
and incorporate the shift in the electric charge of a dyon due to non-zero
expectation values of axion fields [\Wtwo]. Indeed, if the scalars $\phi^i$
tend to constant values $\bar\phi^i$ at spatial infinity, then
$$
q_I = m_{IJ}(\bar\phi) Q^J + a_{IJ}(\bar\phi) p^J\ .
\eqn\aseven
$$
The charges $(p^I, q_I)$ can be combined into a $2k$-vector
$$
{\cal Z} = \oint_\Sigma {\cal F} = \pmatrix{p^I\cr q_I}
\eqn\aeight
$$
from which it is clear that ${\cal Z}\rightarrow \Lambda {\cal Z}$ under
$G$.

The Dirac-Schwinger-Zwanziger (DSZ) quantization condition (with
$\hbar=1$) for two dyons with charge vectors ${\cal Z}$ and ${\cal Z}'$ is
$$
{\cal Z}^T \Omega {\cal Z}' \equiv p^I q_I' -p'{}^I q_I = \nu
\eqn\anine
$$
for some integer $\nu$. This quantization condition is manifestly
$G$-invariant as $G\subseteq Sp(2k;\R)$. However, it has implications for the
quantum theory only if there exist both electric and magnetic charges. If,
for example, there are no magnetic charges of one type then \anine\ places no
constraint on the values of the corresponding electric charge.
For   the
cases of interest to us here, we will show that there exist electric and
magnetic
charges of all types.  We shall now proceed with our analysis of the general
case   assuming all types of charge exist and this, together with the
quantization condition
\anine, implies that the Noether electric charges $q_I$ lie in some lattice
$\Gamma$ and that the magnetic charges $p^I$ lie in the dual lattice $\tilde
\Gamma$. The group $G$ is therefore broken to the discrete sub-group $G(\Z)$
which has the property that a vector ${\cal Z} \in\Gamma\oplus \tilde\Gamma$
is taken to another vector in the same self-dual lattice.  The subgroup of
$Sp(2k)$ preserving the lattice is $Sp(2k;\Z)$, so that the duality group is
$$
G(\Z) = G\cap Sp(2k;\Z)\ .
\eqn\aten
$$
For compact $G$, $G(\Z)$ is a finite group while for non-compact $G$ it is an
infinite discrete group.
If we choose a basis for the fields $A^I$ so that the electric charges, and
hence the magnetic charges, are integers, then the lattice $\Gamma\oplus
\tilde\Gamma$ is preserved by integer-valued matrices, so that $Sp(2k;\Z)$
consists of integer-valued $2k \times 2k$  matrices preserving $ \Omega$, and
$G(\Z)$ is also represented by integer-valued  $2k\times 2k$ matrices.
Note that the group $G(\Z)$ is independent of the geometry of the lattice, as
any two lattices
$\Gamma,
\Gamma'$ are related by a
$GL(k,\R)$ transformation, so that the corresponding discrete groups $G(\Z)$,
$G'(\Z)$ are
related by  $GL(k,\R)$ conjugation and so are isomorphic. For $N=4$
supergravity
coupled to $22$ vector multiplets, $G(\Z)$ is precisely the $S\times T$ duality
group $SL(2;\Z)\times O(6,22;\Z)$ of the toroidally-compactified heterotic
string, which was observed previously to be the quantum symmetry group of this
effective field theory [\Sen]. For $N=8$ supergravity $G(\Z)$ is a discrete
subgroup of $E_{7(7)}$ which we shall call $E_{7(7)}(\Z)$ and abbreviate to
$E_7(\Z)$. It can be alternatively characterized as the subgroup of
$Sp(56;\Z)$ preserving the invariant quartic form of $E_{7(7)}$. {}From the
explicit form of this invariant given in [\CJ], it is straightforward to see
that $E_7(\Z)$ contains an $SL(8,\Z)$ subgroup. We also have
$$
E_7(\Z) \supset SL(2;\Z) \times O(6,6;\Z)\ ,
\eqn\aeleven
$$
so that $E_7(\Z)$ contains the T-duality group of the toroidally-compactified
type II string. The minimal extension of the S-duality conjecture for the
heterotic string would be to suppose that the $SL(2;\Z)$ factor extends to an
S-duality group of the type II string, but it is natural to conjecture that
the full discrete symmetry group is the much larger U-duality group $E_7(\Z)$.
$E_7(\Z) $ is strictly larger than $ SL(2;\Z) \times O(6,6;\Z)$, as it also
contains an $SL(8,\Z)$ subgroup. In the next section we shall verify some
predictions of U-duality for the spectrum of states saturating a gravitational
version of the Bogomolnyi bound, i.e. the `Bogomolnyi states'. However, before
turning to the spectrum we should verify that the Bogomolnyi bound is itself
U-duality invariant, since otherwise a U-duality transformation could take a
state in
the Bogomolnyi spectrum to one that is not in this spectrum.

Consider first the cases of pure $N=4$ supergravity (without matter coupling)
and $N=8$ supergravity, for which $\cM=G/H$ and $k=N(N-1)/2$. We define
$Y_{mn}= t_{mn}^I( \bar q_I +i \bar p^I)$ where $(\bar p^I , \bar q_I)$ are
the components of the $2k$-vector $\bar {\cal Z}= \overline V {\cal Z}$ and
$\overline
V$ is the constant asymptotic value of the $G$-valued field $V$ at spatial
infinity. Here $m,n=1, \dots ,N$ and $ t_{mn}^I=- t_{nm}^I$ are the matrices
generating the vector representation of $SO(N)$. The $Y_{mn}$ appear in a
global supersymmetry algebra as central charges [\WO,\GH,\G,\cmh] and this
allows a derivation of a Bogomolnyi bound. The anti-symmetric complex
$N\times N$ matrix $Y_{mn}$ has $N/2$ complex skew eigen-values $\lambda _a$,
$a=1,\dots, N/2$ and the bound on the ADM mass of the Maxwell-Einstein
theory [\GH] can be generalized to [\cmh]
$$
M_{ADM} \ge max \vert \lambda _a \vert
\eqn\atwelve
$$
Since ${\cal Z}\rightarrow \Lambda{\cal Z}$ and $\overline V \rightarrow
\overline
V\Lambda^{-1}$ under $G$, it follows that $\bar {\cal Z}$ and the $\lambda_a$
are invariant under duality transformations, so the bound \atwelve\ is
manifestly $G$-invariant. In the quantum theory this bound translates
to a bound on the mass of the corresponding quantum state. Similar results
apply to the case of $N=4$ super-matter coupled to supergravity, with the
difference that $t_{mn}^I=- t_{nm}^I$ are now certain scalar-field dependent
matrices that \lq convert' the $SO(6,m)$ index $I$ to the $SO(6)$ composite
index $mn$. Nevertheless, the charges $\lambda_a$ remain duality-invariant.

If  the moduli of all the eigenvalues are equal, $ \vert\lambda_{a_1}\vert=
\vert\lambda_{a_2}\vert=\dots =\vert\lambda_{a_{N/2}}\vert$, then the bound
\atwelve\ is equivalent to
$$
M_{ADM} \ge \sqrt
{ {2 \over N} } \sqrt {\vert {\bar  {\cal Z}} \vert^2}
\eqn\btwelve
$$
where
$$
\vert \bar { {\cal Z} }\vert^2={ \sum_a  { \vert \lambda _a \vert ^2} }
={1 \over 2} \overline Y_{mn}\overline Y^{mn} = G_{IJ}  \bar p^I \bar p^J+
G_{IJ} \bar q_I
\bar q_J
\eqn\ezis
$$
and $G_{IJ}= {1 \over 2}t_{mn}^It_{mn}^J$ is the identity matrix for pure
supergravity, but is scalar dependent for the the matter-coupled $N=4$ theory.
However, in the general case of different eigenvalues, the bound
\ezis\ is strictly weaker than \atwelve. If $M_{ADM}$ is equal to the modulus
of
$r$ of the eigenvalues $\lambda_a$, $M_{ADM}= \vert\lambda_{a_1}\vert=
\vert\lambda_{a_2}\vert=\dots =\vert\lambda_{a_r}\vert$, for some $r$ with $ 0
\le r \le N/2$, then the soliton with these charges spontaneously breaks the
$N$ original supersymmetries down to $r$ supersymmetries, so that for solitons
for which $r=N/2$  precisely half of the $N$ supersymmetries are preserved and
the bound \atwelve\ is equivalent to \ezis. The duality invariance of the
bound   \ezis\ for $N=4$ was previously pointed out in [\SS,\Sen].

%%%%%%%%%%%%%%%%%%%%%%%%%%%%%%%Chapter3%%%%%%%%%%%%%%%%%%%%%%%%%%%%%%%%

\chapter{Spectrum of Bogomolnyi states}

There are many massive states in the spectrum of toroidally-compactified
string theories. The masses of those which do not couple to any of the
$U(1)$ gauge fields cannot be calculated exactly. This is also true in general
of those that do couple to one of the $U(1)$ gauge fields but the masses of
such particles are bounded by their charges, as just described. It is
believed that the masses of string states that saturate the bound
are not renormalized for theories with at least $N=4$ supersymmetry. If this is
so then these masses can be computed exactly. Such `Bogomolnyi states' arise in
the theory from winding and Kaluza-Klein modes of the fundamental string, and
from quantization of non-perturbative soliton solutions of the string theory.
The latter include extreme black holes and, for the heterotic string,
self-gravitating BPS monopoles.

For generic compactifications of both heterotic and type II strings there are
28 abelian gauge fields and so a possible 56 types of electric or magnetic
charge. We  shall identify solitons of the effective supergravity theory
carrying each type of charge, thereby justifying the quantization condition on
these charges. These solitons are various types of extreme black holes.
Initially, at least, we shall be interested in solitons carrying only one type
of charge, in which case we should consistently truncate the supergravity
theory
to one with only one non-zero field strength, $F$. The coefficients of the
$F^2$
terms can then be expressed in terms of a scalar field $\sigma$ and a
pseudoscalar field $\rho$ (which are two functions of the $\phi ^i$) such that
the truncated field theory has an action of the form
$$
S=\int
 d^4x \sqrt
{-g} \left({1\over4} R+ {1 \over 4}e^{-2a \sigma} F_{\mu \nu}F^{\mu\nu}+ {1
\over
4}\rho F_{\mu \nu}{\star F} ^{\mu\nu}
+L(\sigma,\rho)\right)
\eqn\eact
$$
where $L(\sigma,\rho)$ is the lagrangian for a scalar sigma-model and $a$ is a
constant. One can choose $a\ge0$ without loss of generality since $a$ is
changed to $-a$ by the field redefinition $\sigma\rightarrow -\sigma$. For
every
value of $a$ the equations of motion of \eact\ admit extreme multi-black hole
solutions [\GETC], parameterised by the asymptotic values of $\sigma,\rho$,
which are arbitrary integration constants.
There is an intrinsic ambiguity in the metric of the $a\ne 0$ extreme black
hole
solutions because a new metric can be constructed from the canonical metric
(appearing in the action \eact) by a conformal rescaling by a power of
$e^\sigma$. The general metric in this conformal equivalence class will not
have an interpretation as a \lq soliton' in the sense for which the $a=0$
extreme
Reissner-Nordstrom (RN) black hole is a soliton. One feature that is generally
expected from a
soliton is that it interpolate between different vacua: in the RN case these
are the Minkowski spacetime near spatial infinity and the Robinson-Bertotti
vacuum down an infinite Einstein-Rosen `throat'. If we require of the $a\ne0$
extreme black holes that they have a similar property then one must rescale the
canonical metric $d\hat s^2$ by $e^{2a\sigma}$, after which one finds, for
vanishing
asymptotic values of $\sigma$ and $\rho$, the solution
$$
\eqalign{ds^2 &=
e^{2a\sigma}d\hat s^2  \cr
&=
-\Big(1-{(1+a^2)M\over r}\Big)^{2(1-a^2)\over (1+ a^2)} dt^2 +
\Big(1-{(1+a^2)M\over r}\Big)^{-2} dr^2 + r^2 d\Omega_2^2\cr
e^{-a\sigma} &= \Big( 1-{(1+a^2)M\over r}\Big)^{{a^2\over 1+ a^2}},
\qquad \rho=0
\cr}
\eqn\emet
$$
where $M$ is the ADM mass and $d\Omega_2^2$ is the metric on the unit 2-sphere.
 When $a=1$ and $\sigma$ is the dilaton field this rescaling of the canonical
metric is exactly what is required to get the so-called \lq string metric', so
that the $a=1$ black holes have a natural interpretation as string solitons.
This
might make it appear that the rescaling of the canonical metric by
$e^{2a\sigma}$ is inappropriate to string theory when $a\ne 1$, but it must be
remembered that the scalar field $\sigma$ is not necessarily the dilaton but
is, in general, a combination of the dilaton and modulus fields of the torus
and gauge fields.
Indeed, it was shown in [\DKMR] that for the $a=\sqrt{3}$ black holes this
combination is such that the effective rescaling is just that of \emet.
For any value of $a$ this metric has an internal infinity as $r\rightarrow
(1+a^2)M$ for constant $t$. For $a<1$ the surface $r= (1+a^2)M$ is an event
horizon, but this event horizon is regular only if $2(1-a^2)/(1+a^2)$ is an
integer, which restricts the values of $a$ less than unity to $a=0$ or
$a=1/\sqrt{3}$. The $a=0$ case is the extreme RN black hole
for which the soliton interpretation is widely accepted.
The significance of the $a=1/ \sqrt
 3$ case has been explained in [\GHT].
For $a\geq 1$ the surface
$r=(1+a^2)M$ is at infinite affine parameter along any geodesic so one might
admit
all values of $a\ge 1$. On the other hand, the relevance of geodesic
completeness
is not clear in this context so one might still wish to insist that
$2(1-a^2)/(1+a^2)$
be an integer so that the null surface $r= (1+a^2)M$ is regular, in which case
only the further values of $a=1$ and $a=\sqrt{3}$ can
be admitted. Curiously, the values
$$
a=0,\ {1\over \sqrt{3}}, \ 1,\  \sqrt{3}
\eqn\valuea
$$
which we find in this way by demanding that the solution \emet\ is a
{\sl bone fide} soliton also arise from truncation of $N=8$ supergravity. The
possibility of the values $a=0$ and $a=1$ is guaranteed by the existence of
consistent truncations of $N=8$ supergravity to $N=2$ and $N=4$ supergravity
respectively. The possibility of the values $a=\sqrt{3}$ and
$a={1\over\sqrt{3}}$
is guaranteed by the existence of a consistent truncation of the maximal
five-dimensional supergravity to simple five-dimensional supergravity since
the subsequent reduction to four dimensions yields just these values.

Consider first the $a=0$, electric and magnetic extreme RN black holes. Given
any one such black hole with integral charge, an infinite number can be
generated
by acting with $G(\Z)$, and these will include black holes carrying each of the
56
types of charge [\cmh], and this is already sufficient to show that the
continuous duality group $E_{7(7)}$ is broken to a discrete subgroup. These
solutions break $3/4$ of the supersymmetry in the
$N=4$ theories and $7/8$ of the supersymmetry in the $N=8$ case.
For the remainder of the paper, we shall restrict ourselves to
  solitons which break half the supersymmetry,
and the only extreme black hole solutions of this type
are those with $a=\sqrt{3}$. This follows from  consideration of the
implications  of supersymmetry for the moduli space of multi-black hole
solutions. This multi-soliton moduli space is the target space for an effective
sigma
model describing non-relativistic solitons [\Gauntlett]. This sigma model must
have 8 supersymmetries for solitons of a four-dimensional $N=4$ supergravity
theory that break half the supersymmetry, and this implies that the moduli
space is hyper-Kahler. Similarly, the moduli space for multi-solitons of $N=8$
supergravity that break half the supersymmetry is the target space for a sigma
model with 16 supersymmetries, and this implies that the moduli space is flat.
However, the moduli space of multi-black hole solutions  is flat if and only if
$a=\sqrt{3}$ [\Ruback,\Shiraishi], so only these extreme black holes can be
solutions of $N=8$ supergravity that break half the supersymmetry. An
alternative
characterization of these extreme black holes is as `compactifications' of the
extreme black $p$-brane solitons of the ten-dimensional supergravity theory,
which are known to break half the supersymmetry [\DLtwo]. It follows that the
moduli space of these solutions must be flat and what evidence there is [\FS]
confirms this prediction. This ten-dimensional interpretation of the solitons
discussed here will be left to the following section where it will also become
clear that they carry combinations of all $28+28$ electric and magnetic charges
associated with the $28$  {}  $U(1)$ gauge fields.

This moduli space argument shows, incidentally, that whereas the  flatness of
the moduli space for
solitons that break half the supersymmetry is
protected by supersymmetry for $N=8$
supergravity, this is not so    for $N=4$ theories. There is then no reason to
expect the moduli
space metric of extreme black hole solitons of the exact heterotic string
theory (to
all orders in $\alpha'$ and $g$) to be flat. Indeed,  the
$a=\sqrt{3}$ extreme black holes, which have a flat moduli space, are only
approximate solutions of
the heterotic string and are expected to receive higher order corrections.
 Furthermore, if BPS-type monopoles were to occur in the
type II string theory, a possibility that is suggested by the occurrence of
non-abelian gauge groups in some versions of the compactified type II string
[\GD], they would have to break more than half the supersymmetry as their
moduli space is not flat.
This is in accord with the fact that the four-dimensional type II
strings of ref. [\GD] have at most $N=4$ supersymmetry, so that solitons of
these
theories saturating a Bogomolnyi bound would have less than $N=4$
supersymmetry.
This provides further justification for our assumption
that the solitons of the toroidally
compactified type II string that break half the $N=8$ supersymmetry are those
of
the effective $N=8$
supergravity theory.

The complete set of soliton solutions of a supergravity theory fill out
multiplets of the duality group $G(\Z)$, as mentioned in the
introduction. We shall now explain this in more detail. Flat four-dimensional
space-time with the scalar fields $\phi^i$ taking constant values,
$\phi^i_0$, is a vacuum solution of the supergravity theory parameterised by
these constants. The duality group acts non-trivially on such vacua as it
changes the $\phi ^i_0$. The solitons for which the scalar fields tend
asymptotically to the values $\phi^i_0$ provide the solitonic Bogomolnyi states
about the vacuum state $\vert \phi_0>$. A $G(\Z)$ transformation takes a
Bogomolnyi state in this vacuum with charge vector ${\cal Z}$ to another
Bogomolnyi state with charge vector ${\cal Z}'$ and equal mass but in a new
vacuum $\vert \phi'_0>$. As in [\Sen], it will be assumed that one can smoothly
continue from ${\phi '}_0$ to $\phi_0$ without encountering a phase
transition, to obtain a state with the charge vector ${\cal Z}'$, but a
different mass in general, about the original vacuum $\vert \phi_0>$. This
assumption seems reasonable because the extreme black hole solutions depend
analytically on the constants $\phi^i_0$. We thus obtain a new Bogomolnyi
soliton solution about the original vacuum but with a $G(\Z)$ transformed
charge vector. The spectrum of all the Bogomolnyi states obtained in this way
is $G(\Z)$ invariant by construction. In particular, the number of these
Bogomolnyi states with charge vector ${\cal Z}$ will be the same as the number
with charge vector ${\cal Z}'$ whenever ${\cal Z}$ is related to ${\cal Z}'$
by a $G(\Z)$-transformation.

In addition to the Bogomolnyi states that arise from solitons, there are also
the electrically charged Bogomolnyi states of the fundamental string. These
states are purely
perturbative and for the type II string they consist of the Kaluza-Klein (KK)
and winding modes of the string. If they are also to fit into multiplets of the
duality group they must have magnetically charged partners under duality, and
these should be
non-perturbative, i.e. solitonic. The soliton duality multiplets are,
however, already complete for the reason just given. In order to have duality
of the string theory
we must therefore identify the fundamental states with electrically charged
solitonic
states. We shall see in the next section how this identification must be made.

It might be thought that all electrically charged soliton states should have
an equivalent description in terms of fundamental states. This is presumably
true of the heterotic string since there are fundamental string states carrying
each of the 28 types of electric charge and these are related by the T-duality
group $O(6,22;\Z)$. In contrast, the fundamental modes of the type II string
carry only 12 of the possible 28 electric charges, because the 16
Ramond-Ramond (RR) $U(1)$ gauge fields couple to the string through their
field strengths only. The  12 string-mode electric charges are related by the
T-duality group
$O(6,6;\Z)$ of the type II string. It would be consistent with S and T-duality
to suppose that there are no charged states coupling to the 16 (RR) gauge
fields, but this would not be consistent with U-duality, as we now show.

Recall that an $n$-dimensional representation of $G$ gives an action of $G$ on
$\R^n$ which restricts to an action of $G(\Z)$ on the lattice $\Z^n$. For both
the heterotic and type II strings, the charge vector ${\cal Z}$ transforms
under
$G$ as a 56-dimensional representation. For the heterotic string,
$G=SL(2;\R)\times O(6,22)$ and ${\cal Z}$ transforms according to its
irreducible $({\bf 2}, {\bf 28})$ representation. This has the decomposition
$$
({\bf 2}, {\bf 28}) \rightarrow ({\bf 2},{\bf 12}) + 16\times ({\bf 2},{\bf
1})
\eqn\bone
$$
in terms of representations of $SL(2;\R)\times O(6,6)$. This is to be
compared with the type II string for which $G= E_{7(7)}$ and ${\cal Z}$
transforms according to its irreducible {\bf 56} representation, which has
the decomposition
$$
{\bf 56} \rightarrow ({\bf 2},{\bf 12}) + ({\bf 1}, {\bf 32})
\eqn\btwo
$$
under $SL(2;\R)\times O(6,6)$.
In both cases there is a common sector corresponding to the $({\bf 2},{\bf
12})$ representation of $SL(2;\R)\times O(6,6)$, plus an additional
32-dimensional representation corresponding, for the heterotic string, to the
charges for the additional $U(1)^{16}$ gauge group and, for the type II
strings, to the charges for the Ramond-Ramond (RR) sector gauge fields. It is
remarkable that the latter fit into the irreducible spinor representation of
$O(6,6)$. These decompositions of the {\bf 56} representation of $G$ induce
corresponding decompositions of representations of $G(\Z)$ into representations
of $SL(2;\Z)\times O(6,6;\Z)$ on the charge lattice $\Z^{56}$. In particular,
{\it U-duality requires the 16 +16 electric and magnetic charges of the RR
sector to exist and to transform irreducibly under the action of the T-duality
group
$O(6,6;\Z)$}, and we   conclude that all charges in the RR sector
must be carried by solitons. We shall later confirm this.

%%%%%%%%%%%%%%%%%%%%%%%%%%%%%%%Chapter4%%%%%%%%%%%%%%%%%%%%%%%%%%%%%%%%

\chapter{$p$-Brane interpretation of Bogomolnyi solitons}

We have seen that the solitons of toroidally compactified superstrings fit into
representations of the duality group $G(\Z)$. Our concern here will be to
identify states that break half the supersymmetry and carry just one of the 56
types of electric or magnetic charge. We shall call such states for which the
charge takes the minimum value `elementary'; acting on these with the duality
group $G(\Z)$ will generate a lattice of charged states. Here we wish to show
how the elementary solitons arise from extreme black $p$-brane solitons of the
ten dimensional effective supergravity theory. These may be of electric or
magnetic type. Electric $p$-brane solitons give electrically charged solitons
of the four-dimensional dimensionally reduced field theory, while magnetic
ones give magnetic monopoles, provided we use the form of the four-dimensional
supergravity theory that comes directly from dimensional reduction without
performing any duality transformations on the one-form gauge fields (although
we convert two-form gauge fields to scalar fields in the usual way). If we had
chosen a different dual form of action, the solutions would be the same, but
some of the electric charges would be viewed as magnetic ones, and vice versa.
This form of the action is manifestly invariant under  T-duality: for the
heterotic string, the action is the $O(6,22)$-invariant one given in [\maha],
which is related to the one of [\dR] by a  duality transformation, and for the
type II string, it is a new $O(6,6)$-invariant form of the $N=8$ supergravity
action which is  related to the $SL(8,\R)$-invariant Cremmer-Julia action [\CJ]
by a duality transformation.

An extreme  $p$-brane soliton of the ten-dimensional
 low-energy field theory has a
metric of the form [\HS]
$$
ds^2= A(r)[-dt^2+ dx^i dx^i] + B(r) dr^2 + r^2 d \Omega^2_{8-p}
\eqn\epmet
$$
where $x^i$ ($i=1,\dots , p$) are $p$ flat Euclidean dimensions,
$d\Omega^2_{8-p}$ is the metric on an $(8-p)$-sphere, $r$ is a radial
coordinate, $t$ is a time coordinate and $A(r),B(r)$ are two radial functions
that tend to unity as $ r
\rightarrow \infty$.
These solitons couple either to an anti-symmetric tensor gauge field $A_r$ of
rank $r=7-p$, in which case the $p$-brane is magnetically charged and $F=dA$ is
proportional to the $(8-p)$-sphere volume form $\epsilon_{8-p}$, or one of rank
$r=p-3$, in which case the brane is electric and $\star F$ is proportional to
$\epsilon_{8-p}$. In some cases, the $p$-brane solutions will have corrections
of
higher order in $\alpha '$, but some of the solutions correspond to exact
conformal field theories.

We shall be interested in four dimensional solitons obtained by `compactifying'
$p$-brane solitons on the six-torus. Compactification on $T^p$ is
straightforward since one has only to `wrap' the $p$-brane around the
$p$-torus,
which is achieved by making the appropriate identifications of the $x^i$
coordinates. If $p<6$ a soliton in four dimensions can then be found by taking
periodic arrays on $T^{(6-p)}$ and making a periodic identification\foot{
Alternatively, since the solution of extreme black $p$-branes always reduces to
the solution of the Laplace equation in the transverse space, one has only to
solve this equation on $\R^3 \times T^{(6-p)}$ instead of $\R^{(9-p)}$ to
find solitons of the four-dimensional theory.}. For example [\GHL], a
five-brane
can be wrapped around a five-torus in six ways giving rise to six types of
five-dimensional soliton and these yield six types of  black-hole solitons in
four dimensions on taking periodic arrays. Similarly, to `compactify' a 0-brane
(i.e. a 10-dimensional black hole) on a six-torus one first introduces a
6-dimensional periodic array of such black holes and periodically identifies.
Instead of wrapping all $p$ dimensions of a $p$-brane to obtain a point-like
0-brane in 3+1 dimensions, one can wrap $p-q$ dimensions to obtain a q-brane
soliton in 3+1 dimensions; however, in what follows we shall restrict ourselves
to 0-brane solitons in 4 dimensions.

The bosonic sectors of the ten-dimensional effective field theories of the
heterotic and type IIA and type IIB superstrings each include a metric,
$g_{MN}$, antisymmetric tensor gauge field, $b_{MN}$, and a dilaton field
$\Phi$. We shall first discuss this common sector of all three theories and
then turn to the additional sectors characteristic of each theory. We expect
the solutions we describe to be exact solutions of the classical type II
theory,
and their masses to be unrenormalized in the quantum theory, but
for the heterotic string
they are only approximate solutions (to lowest order in $\alpha '$)
of the low energy field theory.

Dimensional reduction of the common $(g,b, \Phi)$ sector on $T^6$ yields 6
Kaluza-Klein   abelian gauge fields ($g_{\mu i}+\cdots$) coming from
$g_{MN}$ and another 6 abelian gauge fields ($b_{\mu i} +\cdots$) coming from
$b_{MN}$. It is straightforward to identify the magnetically-charged
solitons associated with the KK gauge fields. These are the KK monopoles
[\SGP], consisting of the  product of a self-dual Taub-NUT
instanton, with topology $\R^4$, with a  5-torus and a time-like $\R$.
As this is the the product of a five-metric with a five-torus,
this can also be  viewed as a fivebrane solution of the ten-dimensional theory
wrapped around a five-torus.\foot{ For fixed $r,t$, the solution has topology
$S^3\times T^5$, and the $S^3$ can be regarded as a Hopf bundle of $S^1$ over
$S^2$. Thus locally it is $S^2 \times T^6$, so that this solution might also be
thought of as a twisted 6-brane.} There are six types of KK monopoles in four
dimensions, one for each of the six KK gauge fields, because the fivebrane can
be wrapped around the six-torus in six different ways. As four-dimensional
solutions the KK monopoles are extreme black holes with $a=\sqrt{3}$, as
expected from the moduli space argument of the previous section. The elementary
magnetically charged solitons associated with the $b_{\mu i}$ gauge fields can
be identified with the six possible `compactifications' of the extreme
black fivebrane [\DL,\CHS] of the ten dimensional $(g,b, \Phi)$ theory.
We shall refer to these as abelian H-monopoles; they were first given in
[\khuria] and have been discussed further in [\khurib,\GHL,\DLb]. It is
straightforward to check directly that the KK monopoles and the H-monopoles
are indeed related by T-duality, as expected [\fish,\prep]. Note that we have
not included KK modes of the 5-brane, i.e. configurations in which the 5-brane
has momentum in some of the toroidal directions, as these either lead to
extended objects in four dimensions, or to localised solitons that carry more
than one type of charge and so are not elementary.

The KK and abelian H monopoles have electric duals. These electrically
charged solitons have their ten-dimensional origin in the extreme black
string [\DGHR] of the ($g,b,\Phi$) theory, which is dual  [\RNEP,
\RNEPA,\DLPRL] to the extreme black
five-brane. The 6 electric duals to the abelian
H-monopoles are found by wrapping the solitonic string around the 6-torus,
i.e. the 6 winding modes of the solitonic string. The electric duals of the
KK monopoles come from Kaluza Klein modes of the 1-brane, i.e. configurations
in which the solitonic string has momentum in the toroidal directions. They
can be thought of as pp-waves travelling in the compactified directions [\GP].
These $6+6$ elementary electrically charged solitons couple to the $6+6$ KK and
$b_{\mu i}$ gauge fields. They are in one to one correspondence with the KK
(i.e. torus momentum modes) and winding states of the fundamental string
which couple to the same  $12$ gauge fields. This allows us, in principle, to
identify the fundamental string states as soliton states and, as explained
in earlier sections, U-duality of string theory forces us to do so.

Before turning to solitons of the additional sector of each string theory,
we shall first explain here why these field theory solitons are exact
solutions of type II string theory. Type II string theory in a
$(g,b,\Phi)$ background is described by a non-linear sigma-model with (1,1)
world-sheet supersymmetry. The KK monopole background is described by a (4,4)
supersymmetric sigma-model plus a free (1,1) supersymmetric field theory; this
is conformally invariant [\hypkah] and so gives an exact classical solution of
string theory. The pp-wave background is also an exact classical solution
[\hortset], so that the T-duals of these two solutions must be exact classical
solutions too. In contrast, the heterotic string in a $(g,b,\Phi)$ background
is described by a (1,0) supersymmetric sigma-model, and at least some of the
solutions described above only satisfy the field equations to lowest order in
$\alpha'$.
In some cases, as we will describe later,
  these solutions can be modified to obtain
exact classical heterotic string solutions. However, it is not known in general
whether such
backgrounds can be  modified by higher order corrections to give an exact
string solution.

We have now accounted for  $12+12$ of the required $28+28$ types of charge of
all
three ten-dimensional superstring theories. We now consider how the
additional $16+16$ charges arise in each of these three theories, starting with
the type II string. It is known that, after toroidal compactification, the
type IIA and type IIB string theories are equivalent [\ABEQUIV] but it is
instructive to consider both of them. In either case, we showed in the last
section that U-duality requires that the missing $16+16$ types of charge
transform
as the irreducible spinor representation of the T-duality group. Since
T-duality is a perturbative symmetry, if there were electrically charged states
of this type in the fundamental string spectrum, there would also have to be
magnetic ones. However, magnetic charges only occur in the soliton sector, so a
prediction of U-duality is that the corresponding $16$ electric charges are
also to be found in the soliton sector and not, as one might have thought, in
the elementary string spectrum. We shall confirm this.

First we consider the type IIA theory. The ten-dimensional bosonic massless
fields are the ($g,b,\Phi$) fields of the common sector plus a one-form gauge
potential, $A_M$, and a three-form gauge potential, $A_{MNP}$. These extra
fields appear in the RR sector but couple to the string through their field
strengths {\it only}. Upon compactification to four dimensions, $A_M$ gives
one abelian gauge field $A_\mu$ and $A_{MNP}$ gives 15 abelian gauge fields
$A_\mu^{ij}$. These also couple to the string through their field
strengths only and so there are no elementary string excitations that are
electrically-charged with respect to these 16 gauge fields, as expected. The
solitonic $p$-brane solutions of the ten-dimensional field theory involving
$A_M$ or $A_{MNP}$ and breaking only half the supersymmetry consist of a
$0$-brane, i.e. a (ten-dimensional) extreme black hole, a 2-brane (i.e. a
membrane), a 4-brane and a 6-brane. The $0$-brane and the 2-brane are of
electric type. The $0$-brane gives rise to an electrically-charged
four-dimensional black hole in the toroidally-compactified theory by the
procedure of taking periodic arrays of the ten-dimensional solution. The
membrane gives a total of ${6\times 5\over 2}=15$ electric black holes in four
dimensions after `wrapping' it around two directions of the six-torus and then
taking periodic arrays to construct a four-dimensional solution. Similarly,
the magnetic type 4-brane and 6-brane can be wrapped around the six-torus
(introducing periodic arrays where necessary) to give 15+1 magnetically-charged
black holes in four dimensions. We have therefore found a total of 32
additional electric and magnetic charges. Combined with the previous 24 charges
this gives a total of 56 elementary charged states carrying only one type of
charge. From the low-energy field theory we know that these charges
transform according to the ${\bf 56}$ representation of
$E_7$, and that acting on these elementary solitons with $E_7(\Z)$ generates a
56-dimensional
charge lattice. As anticipated, the extra 16+16 electric and magnetic charges
are inert under S-duality but are mixed by the T-duality group $O(6,6;\Z)$. In
addition to the $p$-brane winding modes discussed above, there are also
$p$-brane
momentum modes; however, to give a 0-brane in 4 dimensions, the $p$-brane must
wrap around the torus as well as having internal momentum, so that the
resulting soliton would carry more than one type of charge and so would not be
elementary; nevertheless, these solitons occur in the charge lattices generated
by the elementary solitons.

A similar analysis can be made for the type IIB theory. In this case the
extra massless bosonic fields in the ten dimensional effective field theory
are a scalar, a two-form potential, $A_{MN}$, and a four-form
potential $A_{MNPQ}^{(+)}$ with self-dual five-form field strength. As for the
type IIA string theory, these gauge fields couple through their field
strengths only and so, again, there are no string excitations carrying the new
electric charges. In the solitonic sector of the ten-dimensional field theory
there is a neutral 5-brane, a self-dual 3-brane and a string, in addition to
the string and neutral fivebrane of the ($g,b,\Phi$) sector. The new neutral
fivebrane gives 6 magnetic charges in four dimensions, the self-dual 3-brane
gives 10 electric and 10 magnetic charges and the new string gives six
electric charges. Note that the new solitonic string couples to the 16
$U(1)$ gauge fields coming from $A_{MN}$ and $A_{MNPQ}^{(+)}$. These 16+16
charges couple to $A_{MN}$, while the fundamental string and the solitonic
string of the common sector both couple to $b_{MN}$; thus it may be
consistent to identify the fundamental and common sector solitonic strings,
but the new solitonic string cannot be identified with either.
As for the type IIA string all 56 charges generate the irreducible ${\bf
56}$-dimensional representation of $E_7(\Z)$.

Finally, we turn to the  heterotic string. We have seen that the common sector
solutions of the
low-energy effective supergravity theory include 12 KK and abelian H
monopoles, and their 12 electric duals,
and   under T-duality these must have 16+16 electric and magnetic
black hole partners coupling to the 16 remaining $U(1)$ gauge fields. These
have a ten-dimensional interpretation as the 0-branes and 8-branes of $N=1$
ten-dimensional supergravity coupled to 16 abelian vector multiplets [\HS]
(which can be taken to be those of the $U(1)^{16}$ subgroup of $E_8\times E_8$
or $SO(32)/\Z_2$).

In addition to these black hole solutions, there are also BPS monopole
solutions of the heterotic string arising from
wrapping heterotic or gauge five-branes around the six-torus [\HL]. The BPS
monopoles are not
solutions of the effective supergravity theory with abelian gauge group, but it
 has been argued
(e.g. in [\GHL]) that there should be modifications  of these monopoles that
are
solutions of the
abelian theory. The moduli spaces for multi-soliton solutions of BPS monopoles
are hyper-Kahler [\hypbps]
while those for   extreme $a = \sqrt
 3$  black holes are flat [\Ruback,\Shiraishi], so that the black holes and BPS
monopoles should
not be related by
duality.  If the modified BPS monopoles also have a non-flat moduli space, then
they too cannot be
dual to black holes. However, it is also possible that they   have a flat
moduli space, and even
that they are equivalent to black hole solutions. The modified BPS monopoles,
if they exist,  would
have electric partners under $S$-duality which would be electric solitons.  The
magnetic partners
under S-duality  of electrically charged Bogomolnyi fundamental string states
are expected to be
magnetic monopole solitons, which might be either BPS-type solutions, or black
holes (or both, if
they are equivalent). In either case, the fundamental string states should be
identified with the
electrically charged solitons related to the magnetic monopoles by S-duality.

Whereas the solutions of the type II string we have discussed are exact
conformal field theories,
 the
solutions of the heterotic string are only approximate low-energy solutions.
 However, some of
these heterotic solutions   have a compact holonomy group,  and for these
one can set the Yang-Mills
connection equal to the spin-connection so that the sigma-model becomes one
with (1,1) supersymmetry
and the resulting background is an exact solution of string theory.
 Applying this to the five-brane gives the symmetric five-brane solution
[\khurib]
and
this can also be
used to construct a \lq symmetric KK monopole'.
However, we do not know which of
the other  solutions of the low-energy effective theory
can be corrected to give exact solutions,  and the duality symmetry of
non-abelian phases of the
theory are not understood.
Moreover  the $\alpha'$ corrections to the four-dimensional supergravity action
give a theory
that is not S-duality invariant, but if string-loop corrections are also
included, an S-duality invariant
action should arise.

%%%%%%%%%%%%%%%%%%%%%%%%%%%%%%%%%%%%%
%%%%%%%%%%%%%%%%%%%
%%%%%%%%%%%%%%%%%%%%%%%%%%%%%%%Chapter5%%%%%%%%%%%%%%%%%%%%%%%%%%%%%%%%

\chapter{ Toroidal compactification to other dimensions}

In this section, we extend the previous discussion to consider the duality
symmetries of type II and heterotic strings toroidally compactified to $d$
dimensions. The resulting low-energy field theory is a $d$-dimensional
supergravity theory which has a rigid \lq duality' group $G$, which is a
symmetry of the equations of motion, and in odd dimensions is in fact a
symmetry of the action.  In each case the massless scalar fields of the theory
take values in $G/H$, where $H$ is the maximal compact subgroup of $G$.
$G$ has an $O(10-d,10-d)$ subgroup for the type II
string, and an $O(10-d,26-d)$ subgroup  for the heterotic string. In either
 string theory, it is known that this subgroup is broken down to
the discrete T-duality group,  $O(10-d,10-d;\Z)$ or $O(10-d,26-d;\Z)$.  It
is natural to conjecture that the whole supergravity duality group $G$ is
broken down to a discrete subgroup $G(\Z)$ (defined below) in the
$d$-dimensional string
theory. We have already seen that this occurs for $d=4$ and will argue that for
$d>4$ the symmetry
$G$ is broken to a discrete subgroup by a generalisation of the Dirac
quantization condition. In
tables 1 and 2, we list these   groups for toroidally compactified superstring
theories (at a
generic point in the moduli space so that the gauge group is abelian).

\vskip 1cm

\begintable
$\hbox{ Space-time} \atop \hbox { Dimension d} $ | Supergravity Duality Group
$G$ | String T-duality   | $ \hbox {Conjectured} \atop  \hbox {Full String
Duality}$ \elt
 $10A$ | $SO(1,1)/ \Z_2$ | $\II $ | $ \II $ \elt
 $10B$ | $SL(2,\R)$ | $\II $ | $SL(2,\Z)$ \elt
 $9$ | $SL(2,\R)\times O(1,1)$ | $\Z_2$ | $SL(2,\Z)\times \Z_2$ \elt
 $8$ | $SL(3,\R)\times SL(2,\R)$ |  $O(2,2;\Z) $ | $SL(3,\Z)\times SL(2,\Z)$
\elt
 $7$ | $O(5,5)$ | $O(3,3;\Z)$ | $O(5,5;\Z)$ \elt
 $6$ | $SL(5,\R)$ | $O(4,4;\Z)$ | $SL(5,\Z)$ \elt
 $5$ | $E_{6(6)}$ | $O(5,5;\Z)$ | $E_{6(6)}(\Z)$ \elt
 $4$ | $E_{7(7)}$ | $O(6,6;\Z)$ | $E_{7(7)}(\Z)$ \elt
 $3$ | $E_{8(8)}$ | $O(7,7;\Z)$ | $E_{8(8)}(\Z)$ \elt
 $2$ | $E_{9(9)}$ | $O(8,8;\Z)$ | $E_{9(9)}(\Z)$ \elt
 $1$ | $E_{10(10)}$ | $O(9,9;\Z)$ | $E_{10(10)}(\Z)$
\endtable

\centerline{{\bf Table 1} Duality symmetries for type II string compactified to
$d$ dimensions.}

\vskip .5cm

For the type II string, the supergravity duality groups $G$ are given in
[\julia , \juliab]\foot{They have also been discussed in the context of
worldvolume actions of extended objects in supergravity backgrounds [\DLmem].}.
The Lie algebra of $E_{9(9)}$ is the $E_{8(8)}$ Kac-Moody
algebra,
while  the algebra corresponding to the $E_{10}$   Dynkin diagram has   been
discussed in   [\juliab, \Nic].
The $d=2$ duality symmetry contains the infinite-dimensional Geroch symmetry
group of toroidally
compactified general relativity.
   In $d=9$, the conjectured
duality group is a product of an $SL(2, \Z)$ S-duality and a $\Z_2$
T-duality, while for $d<8$ we conjecture a unified U-duality. For $d=8$, the
T-duality group $O(2,2;\Z) \sim  [SL(2,\Z)\times SL(2,\Z)]/ \Z_2 \times \Z_2$
is a
subgroup of
the conjectured duality. In $d=10$, the type IIA string has   $G=SO(1,1)/
\Z_2$,
while
the type IIB has
$G=SL(2,\R)$, as indicated in the first two lines of the table.
 We shall   abbreviate $E_{n(n)}(\Z)$ to  $E_{n }(\Z)$ when no confusion can
arise.

\vskip 1cm

\begintable
 $\hbox{ Space-time} \atop \hbox { Dimension d}$ | $ \hbox { Supergravity}
\atop
\hbox { Duality Group
$G$ }$ | String T-duality   | $ \hbox {Conjectured} \atop  \hbox {Full String
Duality}$ \elt
 $10$ | $O( 16)\times SO(1,1)$ | $O(16;\Z)$ | $O(16;\Z)\times \Z_2$ \elt
 $9$ | $O(1,17)\times SO(1,1)$ | $O(1,17;\Z)$ | $O(1,17;\Z)\times \Z_2$ \elt
 $8$ | $O(2,18)\times SO(1,1)$ | $O(2,18;\Z)$ | $O(2,18;\Z)\times \Z_2$ \elt
 $7$ | $O(3,19)\times SO(1,1)$ | $O(3,19;\Z)$ | $O(3,19;\Z)\times \Z_2$ \elt
 $6$ | $O(4,20)\times SO(1,1)$ | $O(4,20;\Z)$ | $O(4,20;\Z)\times \Z_2$
\elt
 $5$ | $O(5,21)\times SO(1,1)$ | $O(5,21;\Z)$ | $O(5,21;\Z)\times \Z_2$ \elt
 $4$ | $O(6,22)\times SL(2,\R)$ | $O(6,22;\Z)$ | $O(6,22;\Z)\times SL(2,\Z)$
\elt
 $3$ | $O(8,24)$ | $O(7,23;\Z)$ | $O(8,24;\Z)$ \elt
 $2$ | $O(8,24)^{(1)}$ | $O(8,24;\Z)$ | $O(8,24)^{(1)}(\Z)$ %\elt
% $1$ |  | $O(9,25;\Z)$ |
\endtable

\centerline{{\bf Table 2} Duality symmetries for heterotic string compactified
to $d$ dimensions.}
%\centerline{}

\catcode`\|=12

 \vskip .5cm

For the heterotic string, the supergravity duality groups $G$ for $d>2$ can be
found in articles collected in [\Sezsal]. Pure $N=4$ supergravity in $d=4$
reduces to a
theory with $G=SO(8,2)$ in $d=3$ and to a theory with supergravity duality
group given by  the  affine group $SO(8,2)^{(1)}$ in $d=2$ [\julia]. Similar
arguments
suggest that the heterotic string should give a $d=2$ supergravity theory with
$G$ given by the affine group
$O(8,24)^{(1)}$ symmetry. The heterotic string is conjectured to have
an $S\times T$ duality symmetry in $d \ge 4$ and a unified U-duality in $d \le
3$. Sen  conjectured an $O(8,24;\Z)$ symmetry of $d=3$ heterotic
strings in [\senthree]. The $d=10$ supergravity theory has an $O(16)$ symmetry
acting on the $16$ abelian gauge fields which is broken to the finite group
$O(16;\Z)$; we refer to this as the T-duality symmetry of the ten-dimensional
theory.

The supergravity symmetry group $G$ in $d$ dimensions
doesn't act on the $d$-dimensional space-time
and so survives dimensional reduction.
Then $G$ is necessarily a subgroup of the
symmetry $G'$ in $d'<d$ dimensions and dimensional reduction gives an embedding
of $G$ in $G'$,
and $G(\Z)$ is a subgroup of $G'(\Z)$.
We use this embedding of $G$ into the duality group in $d'=4$ dimensions to
define
the   duality group $G(\Z)$ in $d>4$ dimensions as $G\cap E_7(\Z)$ for the type
II
string  and as $G\cap [O(6,22;\Z)\times SL(2,\Z)]$ for the heterotic string.

The symmetries in $d<4$ dimensions can be understood using a type of argument
first developed to
describe the Geroch symmetry group of general relativity and used in
[\senthree] for $d=3$ heterotic
strings.
The three-dimensional type II string can be regarded
as a four-dimensional theory compactified on a
circle and so is expected to have an $E_7(\Z)$ symmetry. There would then be
seven different $E_7(\Z)$ symmetry groups of the
three dimensional theory corresponding to each of the seven different ways of
first compactifying
from ten to four dimensions, and then from four to three. The seven $E_7(\Z)$
groups and the
$O(7,7;\Z)$ T-duality group  do not commute with each other
and generate a discrete subgroup of $E_8$ which we define to be $E_8(\Z)$.
(Note that the
corresponding Lie algebras, consisting of seven $E_{7(7)}$ algebras and an
$O(7,7)$, generate the
whole of the $E_{8(8)}$ Lie algebra.)
 Similarly, in $d=2$ dimensions, there are eight $E_8(\Z)$ symmetry groups and
an
$O(8,8;\Z)$ T-duality group which generate $E_9(\Z)$ as a  discrete subgroup
of $E_{9(9)}$, and in
the heterotic string there are
eight $O(8,24;\Z)$ symmetry groups from three dimensions and an
$O(8,24;\Z)$ T-duality group which generate $O(8,24;\Z)^{(1)}$ as a  discrete
subgroup of
 $O(8,24)^{(1)}$.

We now turn to the charge quantization condition and soliton spectrum in $d>4$
dimensions. Consider
first the example of type II string theory compactified to $d=5$ dimensions.
The low-energy theory is $d=5$, $N=8$ supergravity [\Cremmer] which has $27$
abelian vector
gauge fields $A^I_\mu$ and an $E_{6(6)}$ rigid symmetry of the action.
Recall that in five dimensions electric charge can be carried by particles or
0-brane solitons,
while magnetic charge can be carried by strings or 1-brane solitons. The 27
types of electric charge
$q_I$ transform as a ${\bf 27}$ of $E_{6(6)}$ while
27 types of magnetic charge
$p^I$ transform as a $\overline {\bf 27}$.
These charges satisfy the quantization condition $q_I p^I= {integer}$
[\RNEP,\Nepteit]
which is invariant under $E_{6(6)}$. As we shall see, all 54 types of charge
occur and so the
electric charges take values in a $27$-dimensional lattice $\Lambda$ and the
magnetic ones take
values in the dual lattice. This breaks the $E_{6(6)}$ symmetry down to the
discrete
 subgroup which preserves the lattice. If the theory is now compactified to
four dimensions,
$E_{6(6)}$ survives as a subgroup of the $E_7{(7)}$ duality symmetry in $d=4$
and the 27-dimensional
lattice
$\Lambda$ survives as a sub-lattice of the 28-dimensional lattice of $d=4$
electric charges (this
will be checked for the elementary charged solitons below).  Thus the subgroup
of $E_{6(6)}$
preserving
$\Lambda$ is
$E_{6(6)}\cap E_7(\Z)$, which is precisely the discrete group $E_6(\Z)$ defined
above.

The five-dimensional theory has a Bogomolnyi bound involving the electric and
magnetic charges   [\GP,\GKLTT]
which is saturated by Bogomolnyi solutions that do not break all the
supersymmetries, and the masses
and charges of these states are expected to be  unrenormalised in the quantum
theory.
 This bound is
invariant under $E_{6(6)}$ and $E_6(\Z)$ and so  Bogomolnyi solitons
automatically fit into
$E_6(\Z)$ representations.

The  $d=5$ elementary solitons of   the type IIA theory carrying precisely one
type of electric or
magnetic charge and breaking half the supersymmetry
 can be identified in a similar manner to that used in $d=4$.
The 27 elementary electrically charged solitons, which are all extreme black
hole solutions in
$d=5$, and the 27 magnetic ones, which are all extreme black strings,
 arise from $d=10$ solutions as follows. The
5-brane wrapped around the 5-torus gives one electrically charged 0-brane and 5
magentically
charged strings. The $d=10$ solitonic string gives 5 electric black holes and 1
black string.
The 0-brane gives one black hole, the 4-brane gives 10 black holes and 5 black
strings, the 4-brane
gives 5 black holes and 10 black strings and the 6-brane gives 1 black string.
In addition, there are 5 electric black holes arising from pp-waves travelling
in each of the 5
toroidal dimensions (these can also be viewed as momentum modes of the
solitonic string), and their
5 magnetic duals, which are a magnetic string generalisation of the KK magnetic
monopole. These are
the solutions consisting of the product of a 4-torus with self-dual Taub-NUT
and two-dimensional
Minkowski space. The non-compact six-dimensional subspace gives rise to a
5-dimensional
magnetic string in the same way that a 4-dimensional KK magnetic monopole
originates from a
five-dimensional solution. Note that these solutions can be thought of as
wrapped 4-brane solutions.
This gives the 27+27 elementary charges, as required.

On further compactification to $d=4$, the 27 electric charges give 27
electrically charged black
holes in $d=4$ and the magnetic strings give 27 magnetic black holes (together
with 27 $d=4$ black
strings). There are two additional elementary charged states in $d=4$, the
pp-wave travelling in
the fifth dimension and the KK monopole corresponding to the fifth dimension;
these two solutions
are uncharged from the five-dimensional point of view.
This corresponds to the fact that the four-dimensional charges lie in a $ {\bf
56 }$ of $E_{7(7)}$
and this decomposes into $E_{6(6)}$ representations as
 $ {\bf 56 } \rightarrow {\bf 27 }+\overline {\bf 27
}+{\bf 1 }+{\bf 1}$.

Similar arguments apply to other strings in
 $d \ge 4$ dimensions, where  charge quantization effects break $G$ to at most
the string
duality groups listed in the tables. In $d$
dimensions there are electric point charges  and  magnetic  $(d-4)$-brane
solitons (which correspond  to a subset of the four-dimensional black hole
solitons on compactification)
and the Dirac quantization of their charges [\RNEP,\Nepteit] breaks the duality
symmetry to the
discrete subgroup $G(\Z)$. There is a similar charge quantization condition on
electric  $p$-branes
and  magnetic $d-p-4$ branes in $d$ dimensions  [\RNEP,\Nepteit] which again
break $G$ to
$G(\Z)$.
In each case, the Bogomolnyi solitons automatically fit into representations of
the duality group
(providing one can continue solutions from one vacuum to another as discussed
in
section 3).

For $d<4$, it is not clear how to understand the breaking of $G$ to $G(\Z)$
directly in terms of
$d$-dimensional quantum effects.
Thus, while we have shown that for $d \ge 4$ the group $G$ is broken to at most
$G(\Z)$, there is
less evidence  for
our conjectures for $d<4$, although
we do know that a subgroup of $G$ is broken to the
discrete T-duality group, and that the solitons will fit into representations
of $G(\Z)$.

%%%%%%%%%%%%%%%%%%%%%%%%%%%%%%%Chapter6%%%%%%%%%%%%%%%%%%%%%%%%%%%%%%%%

\chapter{ Compactification of Type II strings on $K_3\times T^2$}

The analysis of Bogomolnyi states can also be carried out for non-toroidal
compactifications. An interesting example is compactification of the type
II superstring on $K_3\times T^2$ because while $K_3$ has no non-trivial
one-cycles and hence no string winding modes it does have 22 non-trivial
two-cycles around which a $p$-brane for $p>1$ can wrap itself to produce a
($p -2$)-brane which will then produce monopole winding states on $T^2$ if
$p<5$ (taking periodic arrays where necessary). The effective four-dimensional
supergravity can be found by the two stage process of compactification to six
dimensions on $K_3$ [\PKTGSW], followed by a straightforward reduction on
$T^2$. It is an $N=4$ supergravity with an $SL(2;\R)\times O(6,22)$ symmetry
and 28 $U(1)$ gauge fields, exactly as for the compactification of the
heterotic string on $T^6$ at a generic point in the moduli space. In fact, the
four-dimensional supergravity theories are identical because the coupling of
$N=4$ supergravity to $k$ abelian vector multiplets is uniqely determined by
the choice of gauge group [\dR]\foot{Compactification of the heterotic string
on $K_3\times T^2$ leads to a
four-dimensional effective field theory with only $N=2$ supersymmetry, for
which the masses of the Bogomolnyi states might be expected to receive
quantum corrections, so we shall not discuss this case here.}. The analysis
of section 2 again applies, with the result that the duality group is broken
down
to $SL(2;\Z)\times O(6,22;\Z)$ by the charge quantization condition, and the
Bogomolnyi bound is again duality invariant. The soliton spectrum then
automatically fits into representations of $SL(2;\Z)\times  O(6,22;\Z)$ and
the solitons correspond to precisely to the same extreme black holes as were
discussed in section 4 for the heterotic string. However, the ten-dimensional
origin of the elementary charged solutions is now different and we now
discuss these.

Consider first the common $(g,b,\Phi)$ sector. Because $K_3$ has no
isometries and no non-trivial one-cycles all KK modes and string winding modes
arise from the $T^2$ compactification. This yields modes carrying 2+2 types of
electric charge which couple to the 2+2 gauge fields from the metric and
antisymmetric tensor. The corresponding magnetic charges are the KK monopoles
and the H-monopoles. The latter can be interpreted as the winding modes on
$T^2$ of the six-dimensional solitonic string found from `wrapping' the
neutral ten-dimensional fivebrane around the $K_3$ surface [\AETW]. We have
therefore identified the modes carrying just one type of the 4+4 electric and
magnetic charges in this sector. As before, we do not consider modes arising
from the ten-dimensional solitonic string on the grounds that these are not
independent of the fundamental string modes already considered.

Consider now the type IIA string. The additional 24 vector gauge fields
in the four dimensional effective field theory arise from the ten-dimensional
RR gauge fields $A_M$ and $A_{MNP}$. One of these vector gauge fields,
$A_\mu$, is the four-dimensional component of $A_M$. The remaining 23 come
from expressing the ten-dimensional three-form $A_{MNP}$ as the exterior
product of a four-dimensional one-form gauge potential times each of the 22+1
harmonic two-forms of $K_3\times T^2$. We must now find the charged Bogomolnyi
states to which these fields couple. Again we consider states carrying only
one type of charge. The ten-dimensional 0-brane and six-brane solitons
associated with $A_M$ yield, respectively, one electric and one magnetic
four-dimensional black hole coupling to $A_\mu$. The electric 2-brane and the
magnetic 4-brane solitons in ten dimensions produce the Bogomolnyi states
carrying the other 23+23 types of charge coupling to the other 23 gauge
fields. Specifically, the 2-brane can be wrapped around the 22+1 non-trivial
two-cycles of $K_3\times T^2$ to produce 22+1 six-dimensional black holes of
which one can then take periodic arrays to get 22+1 four-dimensional electric
black holes. The four brane can be wrapped around the 22 homology two-cycles
of $K_3$ to give 22 six-dimensional 2-branes, each of which can then be
wrapped around $T^2$ to produce a four-dimensional magnetic black hole.
Alternatively, the four-brane can be wrapped entirely around $K_3$ to give
one six-dimensional black hole which then produces a further magnetic
black hole in four dimensions on taking periodic arrays. We have now found a
total of 24+24 additional electric and magnetic black holes. They each satisfy
the Bogomolnyi bound, because the ten-dimensional $p$-brane solitons do, and
they each carry just one type charge. Combining these with the 4+4 black holes
from the $(g,b,\Phi)$ sector yields a total of 28+28 elementary electric and
magnetic extreme black holes which generate  the $({\bf 2},{\bf 28})$
representation of $SL(2;\Z)\times O(6,22;\Z)$.

Consider instead the type IIB superstring. The RR gauge fields are $A_{MN}$
and $A_{MNPQ}^{(+)}$ which produce 2+22 four dimensional gauge fields upon
compactification on $K_3\times T^2$ [\PKTGSW]. These fields couple to the
soliton states in four dimensions obtained by wrapping the extra solitonic
string and fivebrane, and the self-dual three-brane, around the homology
cycles of $K_3\times T^2$, taking periodic arrays when necessary to get a
four-dimensional soliton (alias extreme black hole). There are two homology
one-cycles and two homology five-cycles so the extra solitonic string and
fivebrane produce 2+2 four-dimensional electric and magnetic black holes.
There are 44 three-cycles so the threebrane produces 44 four-dimensional
solitons. Since the threebrane is self-dual 22 of these are electric and 22
magnetic. Again we have a total of 24+24 additional charges. Combining these
with the 4+4 black holes from the $(g,b,\Phi)$ sector again yields a total of
28+28 elementary electric and magnetic extreme black holes which generate the
$({\bf 2},{\bf 28})$ representation of $SL(2;\Z)\times O(6,22;\Z)$.

Since the type II string compactified on $K_3\times T^2$ and the generic
toroidal compactification of the heterotic string have exactly the same
four-dimensional low-energy field theory it is natural to conjecture that they
might be equivalent string theories. If this is so then the Bogomolnyi
states of the heterotic string discussed at the conclusion of the previous
section would have a straightforward ten-dimensional interpretation after
all. It would have some other remarkable consequences. For example, at special
points of the heterotic string moduli space, there are extra massless fields
and an enhanced (Yang-Mills) symmetry due to non-perturbative
world-sheet effects (i.e. non-perturbative in $\alpha '$, but
perturbative in $g$). If the compactified type II string is equivalent, it
must have the same enhanced symmetry in vacua corresponding to the same
points in the scalar field coset space. This presumably does not arise from
non-perturbative world-sheet effects, so would have to come from
non-perturbative stringy effects, or from Wilson lines and their $p$-brane
generalisations. This would mean that the sigma-model
coupling constant $\alpha '$ of the heterotic string becomes one of the
stringy coupling constants of the type II theory, as might have been expected
from the fact that for the toridally compactified type II string, all
coupling constants are on an equal footing and are mixed up under U-duality.

%%%%%%%%%%%%%%%%%%%%%%%%%%%%%%%Chapter6%%%%%%%%%%%%%%%%%%%%%%%%%%%%%%%%

\chapter{ U-duality and the 11-dimensional supermembrane}

As we have seen, the $E_7(\Z)$ invariance of the spectrum
of soliton states of $N=8$ supergravity is an automatic consequence of the
$E_7(\Z)$ invariance of the equations of motion. The non-trivial features
are, firstly, that these states have an interpretation in terms of
ten-dimensional KK solitons and solitonic $p$-branes and, secondly, that if the
ten-dimensional field theory is considered to be the effective field theory of
the type II string theory then $E_7(\Z)$ invariance requires an identification
of the solitonic string with the fundamental string. $N=8$ supergravity can
also
be obtained by dimensional reduction of 11-dimensional supergravity on $T^7$.
We shall now show that the elementary soliton states of $N=8$ supergravity also
have an interpretation in terms of KK solitons and solitonic $p$-branes of
11-dimensional supergravity. There are 7 KK magnetic monopoles and 7 electric
duals, which are pp-waves of 11-dimensional supergravity [\cmhpp] travelling
in the internal dimensions. The
11-dimensional solitonic  $p$-branes are the electric membrane and the magnetic
fivebrane. Each can be wrapped around the seven-torus to produce 21
four-dimensional solitons. Thus we have a total of 28 electric and 28 magnetic
four-dimensional solitons each carrying one of the 56 types of charge which
are a basis for the irreducible {\bf 56} repesentation of $E_7(\Z)$.
If we now wish to interpret 11-dimensional supergravity as an effective
field theory of a fundamental $E_7(\Z)$ supermembrane theory then we must
identify the fundamental membrane with the solitonic one, just as we were
forced to identify the fundamental string with the (appropriate) solitonic
string.

Consider now the compactification to $\cM_4$ on $K_3\times T^3$ of
eleven-dimensional supergravity [\DNP]. The effective field theory is the
same as that of the type II superstring compactified on $K_3\times T^2$, i.e.
an $N=4$ supergravity with 28 $U(1)$ gauge fields and an $SL(2;\Z)\times
O(6,22;\Z)$ symmetry. The soliton spectrum is also the same.
In the monopole sector we have, firstly, 3 KK monopoles from the $T^3$
factor and, secondly, a further 25 monopoles from wrapping the solitonic
fivebrane around the 3+22 homology five-cycles of $K_3\times T^2$. This gives
a total of 28 monopoles. The 28 electrically charged solitons are the electric
duals of these monopoles which can be understood in terms of the KK and
winding modes of either the solitonic membrane or a fundamental membrane. The
entire set of 56 states can be assigned to the (2,28) representation of
$SL(2;\R)\times O(6,22)$, inducing a corresponding representation of
$SL(2;\Z)\times O(6,22;\Z)$.

These results are encouraging signs that it may be possible to define the
quantum supermembrane theory entirely in terms of the solitonic membrane
solution of eleven-dimensional supergravity. Alternatively, one can envisage a
dual formulation in terms of a fundamental $11$-dimensional
the superfivebrane, in which case the solitonic fivebrane might
be identified with a fundamental fivebrane.
Of possible relevance in this
connection is the fact that the membrane and fivebrane solitons have a
very different global structure. Both have a degenerate Killing horizon but
whereas the membrane horizon conceals a singularity in an interior region
[\DGT], much like the extreme Reissner Nordstrom solution of four-dimensional
Maxwell-Einstein theory, the fivebrane is completely non-singular [\GHT].

%%%%%%%%%%%%%%%%%%%%%%%%%%%%%%%Chapter7%%%%%%%%%%%%%%%%%%%%%%%%%%%%%%%%

\chapter{Comments}

The equations of motion of four-dimensional effective supergravity theories
of compactified superstring theories are invariant under a continuous duality
group $G$ that is broken by quantum effects to a discrete subgroup
$G(\Z)$. For the toroidally-compactified heterotic string at a generic point
in the moduli space, and for the $K_3\times T^2$-compactified type II
superstrings, this group is the $S\times T$ duality group $SL(2;\Z)\times
O(6,22;\Z)$. For the toroidally-compactified type II superstrings it is the
$U$ duality group $E_7(\Z)$ which contains the $S\times T$ duality group
$SL(2;\Z)\times O(6,22;Z)$. Whereas T-duality is known to be an exact
symmetry of string theory at each order in the string coupling constant $g$,
the conjectured S- and T- dualities are non-perturbative. We have provided
evidence for U-duality of the type II string by considering those features of
string theory that are expected to be given exactly by a semi-classical
analysis although it should be emphasized that this evidence depends only
on the form of the effective supergravity theory and would apply equally
to any consistent quantum theory of gravity for which this is the effective
low-energy action. Nevertheless, by supposing this consistent quantum theory
to be string theory our arguments have led us to the remarkable conclusions
that it is necessary to identify certain states of the string with extreme
black holes, and the fundamental string with a solitonic string. We have also
seen that the elementary Bogomolnyi states are extreme black hole solutions
of the low-energy theory, and have shown how these arise from
$p$-brane solitons of the ten-dimensional theory.

The zero modes of   the scalar fields of the low energy field
theory are all coupling constants of the string theory, so that $G(\Z)$
symmetry relates different
regimes in the   perturbation theory in these coupling constants, interchanging
strong and weak
coupling and, in the case of $U$-duality, interchanging $g $ with $\alpha '$,
in the sense of
mixing the quantum loop expansion in
 $g$ with the sigma-model perturbation expansion in $\alpha '$ and the moduli
of the
compactification space.
 For the
compactified type II superstring, any physical quantity (e.g. the S-matrix) can
be expanded in terms of
 the 70 coupling constants associated with zero modes of the 70 scalars.   In
the world-sheet approach to string
theory, one first integrates over the sigma-model degrees of freedom on a
Riemann surface of fixed
genus, obtaining a result parameterised by the sigma-model coupling constants,
and then sums over
genus. As U-duality mixes up all 70 coupling constants, the final result may be
expected to depend
on all 70 scalars in a symmetric way, even though the calculation was very
asymmetrical and in
particular picked out the dilaton to play a special role.
This would hugely constrain the theory and the
assumption of
U-duality, together with $N=8$ supersymmety,  gives us a great deal of
non-perturbative information, and might even
enable us to solve the theory! This structure also suggests that there might be
a new formulation of
string theory which treats all the coupling constants on an equal footing.

We have seen that there are duality symmetries in all dimensions $d \le10$ and
it is interesting to ask whether the  $d<10$ symmetries  can correspond to
symmetries of the 10-dimensional theory.
If e.g. the four-dimensional dualities correspond to  symmetries of the full
ten-dimensional
theory, these  symmetries must interchange   the various $p$-brane  solitons of
the
string-theory; such   symmetries would probably have to be non-local.
If the theories in $d<4$
dimensions  really do have the duality symmetries suggested in section 5, and
if these have analogues in
higher dimensions, this would have remarkable consequences for string theory.
For example, the
U-duality of the three-dimensional heterotic string includes transformations
that would mix the
string with    5-brane solitons in  ten dimensions,
and so would contain the
transformations described as the `duality of dualities' in [\SS]. For the
type II string theory, this
suggests that
$E_{10}(\Z)$ might be a discrete non-local symmetry of the ten-dimensional
string theory!

One of the predictions of S-duality for the heterotic string  is the
presence of certain dyon bound states in the Bogomolnyi spectrum, which
can be translated into to a prediction concerning harmonic forms on the
multi-monopole moduli space [\SenSegal]. It would be interesting to consider
the corresponding predictions for the type II string. Recently, some
strong-coupling evidence for
S-duality  of the heterotic string has been found by studying partition
functions of certain topological field theories [\VW], and again it would be
interesting to seek similar strong-coupling tests for U-duality.

Finally, since $N=8$ supergravity and its soliton spectrum are U-duality
invariant, many of the properties previously thought to be unique to string
theory are in fact already properties of the effective supergravity theory once
account is taken of all soliton solutions. Since only stable states can
appear in an exact S-matrix it is possible that the only states of the exact
toroidally compactified type II string theory are the Bogomolnyi states and
that these are in one to one correspondence with soliton states of the
supergravity theory. This would support a previous suggestion [\PKT] that a
fundamental superstring theory might actually be equivalent to its effective
field theory once solitons of the latter are taken into account.
To pursue this further one would need
to find higher spin ($>2$) soliton states corresponding to the higher spin
states
of string theory. There seems to be no problem in principle with the existence
of such higher-spin  soliton states in $N=8$ supergravity because the $a=0$ and
$a=1$extreme black holes must belong to massive supermultiplets with spins $>2$
as they break more than half the supersymmetry.
 In
this connection it is worth recalling the similarity of the mass/spin relation
for Regge trajectories in four-dimensional string theory, $M^2 \propto {J\over
\alpha'}\, $, with that of the degenerate Kerr solutions of general relativity,
$M^2\propto {J\over G}\, $.

%%%%%%%%%%%%%%%%%%%%%%%%%%%%%%%Acknowledgements%%%%%%%%%%%%%%%%%%%%%%%%%%
%%%%%%
\vskip 0.5cm
\noindent{\bf Acknowledgements}: We are grateful to Louise Dolan, Michael Duff,
Jos\' e Figueroa-O'Farrill,
Jerome Gauntlett, Gary Gibbons, Michael Green,  Bernard Julia, Wafic Sabra,
Trevor Samols, Kellogg Stelle and Edward Witten for helpful comments.

%%%%%%%%%%%%%%%%%%%%%%%%%%%%%%%%%%%%%%%%%%%%%%%%%%%%%%%%%%%%%%%%%%%%%%%%%
%%%%%%

\refout

\end